\documentclass[twocolumn,amsmath,amssymb,aps,footinbib]{revtex4-2}  
\usepackage[dvipdfmx]{graphicx} 
\usepackage{dcolumn}
\usepackage{color}
\usepackage{bm}      
\usepackage{xspace}

\newcommand{\be}{\begin{equation}}
\newcommand{\ee}{\end{equation}}
\newcommand{\bea}{\begin{eqnarray}}
\newcommand{\eea}{\end{eqnarray}}
\newcommand{\beaa}{\begin{eqnarray*}}
\newcommand{\eeaa}{\end{eqnarray*}}

\begin{document}

\title{
  Static and Dynamical Spin Correlations
  in the Kitaev Model at Finite Temperatures
  via Green's Function Equation of Motion
}

\author{Hibiki Takegami}
\email{takegami.hibiki.64h@st.kyoto-u.ac.jp}
 \affiliation{
   Course of Studies on Materials Science, 
  Graduate School of Human and Environmental Studies, 
  Kyoto University, Kyoto 606-8501, Japan
 }

\author{Takao Morinari}
 \email{morinari.takao.5s@kyoto-u.ac.jp}
 \affiliation{
   Course of Studies on Materials Science, 
  Graduate School of Human and Environmental Studies, 
  Kyoto University, Kyoto 606-8501, Japan
 }

\date{\today}

\begin{abstract}
The Kitaev model, renowned for its exact solvability and potential to host non-Abelian anyons, remains a focal point in the study of quantum spin liquids and topological phases. While much of the existing literature has employed Majorana fermion techniques to analyze the model, particularly at zero temperature, its finite-temperature behavior has been less thoroughly explored via alternative methods. In this paper, we investigate the finite-temperature properties of the Kitaev model using the spin Green's function formalism. This approach provides a unified framework for computing key physical quantities, such as spin correlations, magnetic susceptibility, and the dynamical spin structure factor, offering valuable insights into the system's thermal dynamics. 
To solve the equation of motion for the spin Green's function, we truncate the hierarchy of multi-spin Green's functions using the Tyablikov decoupling approximation, which is particularly accurate at high temperatures. Our results show several similarities with Majorana-based numerical simulations, though notable differences emerge. Specifically, both static and dynamical spin-spin correlation functions capture not only $\mathbb{Z}_2$ flux excitations but also simple spin-flip excitations, with the latter dominating the response. Additionally, without explicitly assuming fractionalization, our results for the spin susceptibility and spin relaxation rate suggest the presence of fermionic degrees of freedom at low temperatures.
Unlike Majorana-based approaches, which rely on exact solvability and are inherently limited to the pure Kitaev model, the spin Green's function formalism is well-suited for studying more realistic systems. It can naturally incorporate the effects of magnetic fields, including linear-order terms, and non-Kitaev interactions, making it applicable to a broader range of materials and experimental conditions. By providing an alternative method for analyzing the finite-temperature properties of the Kitaev model, this study complements existing approaches and lays the groundwork for future investigations into real materials and extended models.
\end{abstract}

\maketitle

\section{Introduction}
The Kitaev model \cite{Kitaev2006}, a paradigmatic example of exactly solvable quantum spin liquid systems, has garnered significant attention in the field of condensed matter physics due to its potential to host non-Abelian anyons and topologically protected quantum states \cite{Kitaev2003,Nayak2008}. Defined on a two-dimensional honeycomb lattice, the Kitaev model serves as a cornerstone for exploring the interplay between quantum spin liquids and topological phases of matter \cite{Balents2010}. Not only is the Kitaev model analytically tractable, but it also holds promise for real-world applications in materials with strong spin-orbit coupling \cite{Jackeli2009,Takagi2019}.

One of the most compelling approaches to studying the Kitaev model is through the formalism of Majorana fermions \cite{Kitaev2006,Feng2007,Chen2007,Chen2008a}, which maps spin degrees of freedom onto itinerant fermionic ones, enabling an exact solution at zero temperature. At finite temperatures, however, an exact solution becomes elusive even within the Majorana framework. In this regime, the Kitaev Hamiltonian can be interpreted as a system of itinerant Majorana fermions interacting with fluctuating $\mathbb{Z}_2$ gauge fields, where the $\mathbb{Z}_2$ fields commute with the Hamiltonian. As a result, the system behaves as if it consists of free Majorana fermions under the influence of thermally fluctuating $\mathbb{Z}_2$ gauge fields. Leveraging this picture, previous studies have successfully explored thermodynamic quantities \cite{Nasu2014,Nasu2015} and dynamical spin correlations at finite temperatures \cite{Yoshitake2016,Yoshitake2017,Yoshitake2017b,Motome2020}.

  Despite these successes, Majorana-based approaches have significant limitations. First, the computation of dynamical quantities within this framework requires combining multiple numerical techniques, such as Monte Carlo simulations for thermodynamic quantities\cite{Nasu2014} and cluster dynamical mean-field theory for spin susceptibility\cite{Yoshitake2016,Yoshitake2017}. This introduces computational complexity and fragmentation. Furthermore, to fully capture dynamical spin correlations at low temperatures, additional methods, such as continuous-time quantum Monte Carlo techniques\cite{Yoshitake2017b}, are needed to solve impurity Anderson-like problems\cite{Motome2020}, which further complicates the analysis. Second, Majorana-based approaches are inherently restricted to the pure Kitaev model. In the presence of magnetic fields, they can only accommodate terms up to third order in the field to retain exact solvability, while linear-order terms break this solvability. Moreover, the inclusion of non-Kitaev interactions renders the model no longer exactly solvable, limiting the applicability of these approaches to realistic materials with perturbations beyond the ideal Kitaev Hamiltonian.

  In contrast, the spin Green's function formalism\cite{Tyablikov1962,Kondo1972,Shimahara1991,Winterfeldt1997,Bernhard2002,Schmalfuss2006,Froebrich2006,Vladimirov2017,Morinari2018,Sasamoto2024} offers a unified and versatile approach to studying the Kitaev model and its extensions. This method directly computes the spin Green's function and encapsulates the dynamical properties of spin excitations within the system. Unlike Majorana-based approaches, it provides a framework for calculating both thermodynamic and dynamical quantities without requiring a combination of distinct numerical techniques. Additionally, the spin Green's function formalism is well-suited to incorporating magnetic field effects (including linear-order terms) and non-Kitaev interactions, making it a promising tool for studying real materials that deviate from the pure Kitaev model.

At finite temperatures, the spin Green's function formalism enables the computation of key physical quantities, such as spin correlations, magnetic susceptibility, and the dynamical spin structure factor. However, solving the equation of motion for the spin Green's function introduces a hierarchical structure: for instance, in the Heisenberg model, two-spin Green's functions depend on three-spin Green's functions, which in turn require four-spin Green's functions. To address this complexity, the hierarchical structure is typically truncated using the Tyablikov decoupling approximation\cite{Tyablikov1962}, which represents four-spin Green's functions as products of two-spin correlations and two-spin Green's functions. This approximation is particularly accurate at high temperatures, as demonstrated by the agreement between the spin Green's function approach and high-temperature expansion results \cite{Kondo1972}.  

While more sophisticated Green's function approaches may be necessary for systems with long-range spin-spin correlations, the Kitaev model presents a unique advantage: in the ground state, the spin-spin correlation function is finite only between nearest-neighbor sites \cite{Baskaran2007}. This strongly suggests that the spin-spin correlation function remains short-ranged even at finite temperatures, making the spin Green's function formalism an effective and well-suited tool for analyzing the Kitaev model's thermal properties.

In this paper, we investigate the finite-temperature properties of the Kitaev model using the spin Green's function formalism, with the goal of advancing the understanding of this model's physical implications. By employing this framework, we complement existing studies that primarily rely on Majorana fermion techniques, providing a spin-based perspective to further enrich the analysis. Our findings are systematically compared with Majorana-based numerical simulations, as thoroughly reviewed in Ref.~\onlinecite{Motome2020}, establishing a solid benchmark for evaluating the accuracy and utility of our approach. This detailed comparison highlights both the strengths and the unique insights offered by the spin Green's function method, contributing to a broader and more comprehensive understanding of the Kitaev model.

The remainder of the paper is organized as follows.
In Sec.~\ref{sec:formalism}, we introduce the equation of motion approach
for the spin Green's function and apply it to the Kitaev model.
We first discuss the decoupling procedure,
followed by the derivation of the self-consistent equation,
and then present the formulas for calculating various physical quantities.
Due to specific features of the Kitaev model,
some Green's functions and correlation functions are identically zero,
a point elaborated in Appendix~\ref{app:exact-results}.
In Sec.~\ref{sec:corr-func-T0}, we compute the zero-temperature correlation functions
and compare them with the exact solutions obtained via the Majorana fermion approach.
Finite-temperature results are presented in Sec.~\ref{sec:finiteT},
where we examine the temperature dependence of the internal energy,
correlation functions, excitation energies, and spin susceptibility.
Additionally, we compute the dynamical structure factor
and the spin relaxation rate.
Finally, Sec.~\ref{sec:summary} provides a summary of the results.

\section{Spin Green's Function Formalism}
\label{sec:formalism}
In this section, we develop the formalism used to investigate the finite-temperature properties of the Kitaev model through the spin Green's function approach. We begin by introducing the fundamental concepts of the spin Green's function formalism and then proceed to discuss its specific application to the Kitaev model, including the use of the decoupling approximation to manage the hierarchical structure of the equations of motion.

\subsection{Definition of the Green's function}
We consider a spin-1/2 system described by the Hamiltonian ${\cal H}$,
with the spin at site $i$ represented by
\be
S_i^\mu  = \frac{1}{2}\sigma _i^\mu,
\ee
where $\sigma_i^\mu$ refers to the $\mu$ ($\mu = x, y, z$) component of the Pauli matrices. To describe the system's dynamics, we introduce the Matsubara Green's function. Let $\tau$ denote the imaginary time, and $T_\tau$ the imaginary-time ordering operator. The Matsubara Green's function is then defined as:
\be
G_{ij}^{\mu \nu }\left( \tau  \right)
=  - \left\langle {{T_\tau }S_i^\mu \left( \tau  \right)
  S_j^\nu \left( 0 \right)} \right\rangle
\equiv
    {\left\langle {{S_i^\mu }}
      \mathrel{\left | {\vphantom {{S_i^\mu } {S_j^\nu }}}
        \right. \kern-\nulldelimiterspace}
              {{S_j^\nu }} \right\rangle _\tau },
    \label{eq:Gij}
\ee
where
\be
S_i^\mu (\tau) = e^{\tau {\cal H}} S_i^\mu e^{-\tau {\cal H}}
\ee
is the spin operator in the Heisenberg picture with imaginary time evolution.
This Green's function encapsulates the spin-spin correlations and dynamical properties of the system.
For any operator $A$, its imaginary-time evolution is defined by
\be
A\left( \tau  \right) = {e^{\tau {\cal H}}}A{e^{ - \tau {\cal H}}}.
\ee

If we are able to compute $G_{ij}^{\mu \nu }(\tau)$, it is straightforward to compute various physical quantities, such as the two-spin correlation functions, spin susceptibility, and the dynamical spin structure factor. Additionally, the internal energy can be computed if the Hamiltonian, ${\cal H}$, consists only of two-spin interactions. A familiar approach to finding the expression for $G_{ij}^{\mu \nu }(\tau)$ is through perturbative calculations. However, such an approach is limited to cases where ${\cal H}$ can be divided as ${\cal H} = {\cal H}_0 + {\cal H}_1$, with ${\cal H}_0$ representing a solvable system and ${\cal H}_1$ treated as a perturbation.
This scenario is not generally applicable to most spin systems, such as the Heisenberg model,
the Kitaev model, and others.

In these cases, we need to compute $G_{ij}^{\mu \nu }(\tau)$ using the full Hamiltonian, ${\cal H}$, necessitating a different approach. Here, we adopt the equation of motion method for $G_{ij}^{\mu \nu }(\tau)$. This approach is non-perturbative and does not rely on assumptions about the correlations in the system. Furthermore, as we shall see, the temperature dependence of various physical quantities can be computed, including dynamical quantities, where the analytic continuation is performed exactly.

\subsection{Equation of motion and decoupling approximation}
From Eq.~(\ref{eq:Gij}), $G_{ij}^{\mu \nu }(\tau)$ is expressed as 
\be
G_{ij}^{\mu \nu }\left( \tau  \right)
=  - \theta \left( \tau  \right)\left\langle
{S_i^\mu \left( \tau  \right)S_j^\nu \left( 0 \right)} \right\rangle
- \theta \left( { - \tau } \right)\left\langle {S_j^\nu
  \left( 0 \right)S_i^\mu \left( \tau  \right)} \right\rangle,
\ee
where $\theta(\tau)$ is the step function. Taking the derivative with respect to $\tau$, we obtain
\be
   {\partial _\tau }G_{ij}^{\mu \nu }\left( \tau  \right)
   =  - {\left\langle {{\left[ {S_i^\mu ,{\cal H}} \right]}}
 \mathrel{\left | {\vphantom {{\left[ {S_i^\mu ,{\cal H}} \right]} {S_j^\nu }}}
 \right. \kern-\nulldelimiterspace}
 {{S_j^\nu }} \right\rangle _\tau } - \delta \left( \tau  \right)\left\langle {\left[ {S_i^\mu ,S_j^\nu } \right]} \right\rangle 
 \label{eq:EOM_G_tau}
 \ee
where we used the relation
 \be
    {\partial _\tau }S_i^\mu \left( \tau  \right)
    =  - \left[ {S_i^\mu \left( \tau  \right),{\cal H}} \right].
    \ee
Denoting the inverse temperature as $\beta = 1/T$, where $T$ is the temperature and the Boltzmann constant is set to $k_{\rm B} = 1$, we define the Fourier transform as
    \be
    G_{ij}^{\mu \nu }\left( {i{\omega _n}} \right) = \int_0^\beta  {d\tau } {e^{i{\omega _n}\tau }}G_{ij}^{\mu \nu }\left( \tau  \right)
    \equiv {\left\langle {{S_i^\mu }}
 \mathrel{\left | {\vphantom {{S_i^\mu } {S_j^\nu }}}
 \right. \kern-\nulldelimiterspace}
 {{S_j^\nu }} \right\rangle _{i{\omega _n}}}.
    \ee
    Here, $\omega_n = 2\pi n/\beta$ with $n$ being an integer is the bosonic Matsubara frequency.
    The notation
    ${\left\langle {{S_i^\mu }}
 \mathrel{\left | {\vphantom {{S_i^\mu } {S_j^\nu }}}
 \right. \kern-\nulldelimiterspace}
         {{S_j^\nu }} \right\rangle _{i{\omega _n}}}$
    is introduced for conciseness, and this convention is also applied to similar expressions.
    From Eq.~(\ref{eq:EOM_G_tau}), we derive
    \be
    i{\omega _n}{\left\langle {{S_i^\mu }}
 \mathrel{\left | {\vphantom {{S_i^\mu } {S_j^\nu }}}
 \right. \kern-\nulldelimiterspace}
 {{S_j^\nu }} \right\rangle _{i{\omega _n}}} = {\left\langle {{\left[ {S_i^\mu ,{\cal H}} \right]}}
 \mathrel{\left | {\vphantom {{\left[ {S_i^\mu ,{\cal H}} \right]} {S_j^\nu }}}
 \right. \kern-\nulldelimiterspace}
         {{S_j^\nu }} \right\rangle _{i{\omega _n}}} + \left\langle {\left[ {S_i^\mu ,S_j^\nu } \right]} \right\rangle.
    \label{eq:G_EOM1}
    \ee
    which we refer to as the first-order equation of motion.
    
    In this formalism, the order of an equation of motion corresponds
    to the number of sequential commutation operations performed
    to derive the equation of motion for the Green's function.
    To find
    ${\left\langle {{S_i^\mu }}
 \mathrel{\left | {\vphantom {{S_i^\mu } {S_j^\nu }}}
 \right. \kern-\nulldelimiterspace}
         {{S_j^\nu }} \right\rangle _{i{\omega _n}}}$
    we require
    ${\left\langle {{\left[ {S_i^\mu ,{\cal H}} \right]}}
 \mathrel{\left | {\vphantom {{\left[ {S_i^\mu ,{\cal H}} \right]} {S_j^\nu }}}
 \right. \kern-\nulldelimiterspace}
         {{S_j^\nu }} \right\rangle _{i{\omega _n}}}$.
    In order to find
        ${\left\langle {{\left[ {S_i^\mu ,{\cal H}} \right]}}
 \mathrel{\left | {\vphantom {{\left[ {S_i^\mu ,{\cal H}} \right]} {S_j^\nu }}}
 \right. \kern-\nulldelimiterspace}
         {{S_j^\nu }} \right\rangle _{i{\omega _n}}}$,
    we need
    ${\left\langle {{\left[ {\left[ {S_i^\mu ,{\cal H}} \right],{\cal H}} \right]}}
      \mathrel{\left | {\vphantom {{\left[ {\left[ {S_i^\mu ,{\cal H}} \right],{\cal H}} \right]}
            {S_j^\nu }}}
 \right. \kern-\nulldelimiterspace}
              {{S_j^\nu }} \right\rangle _{i{\omega _n}}}$.
    The second-order equation of motion is obtained by applying a second commutation operation to the Green's function resulting from the first-order equation. This typically introduces Green's functions with more complex spin operator combinations, reflecting the additional layers of dynamics introduced by successive commutations.

In general, this process results in an infinite chain of equations, which necessitates an approximation to truncate the hierarchy. Here, we employ the Tyablikov decoupling approximation \cite{Tyablikov1962}, in which the Green's function involving $n$ spins is approximated as the product of a spin correlation function with $m$ spins and a Green's function with $n-m$ spins. In this decoupling process, we introduce a parameter to enhance the accuracy of the approximation. The number of parameters introduced is constrained by the number of self-consistent equations available. This point will be further clarified below through the explicit application of the formalism to the Kitaev model.
    
\subsection{Application to the Kitaev model}
\begin{figure}[htbp]
  \includegraphics[width=0.8 \linewidth, angle=0]{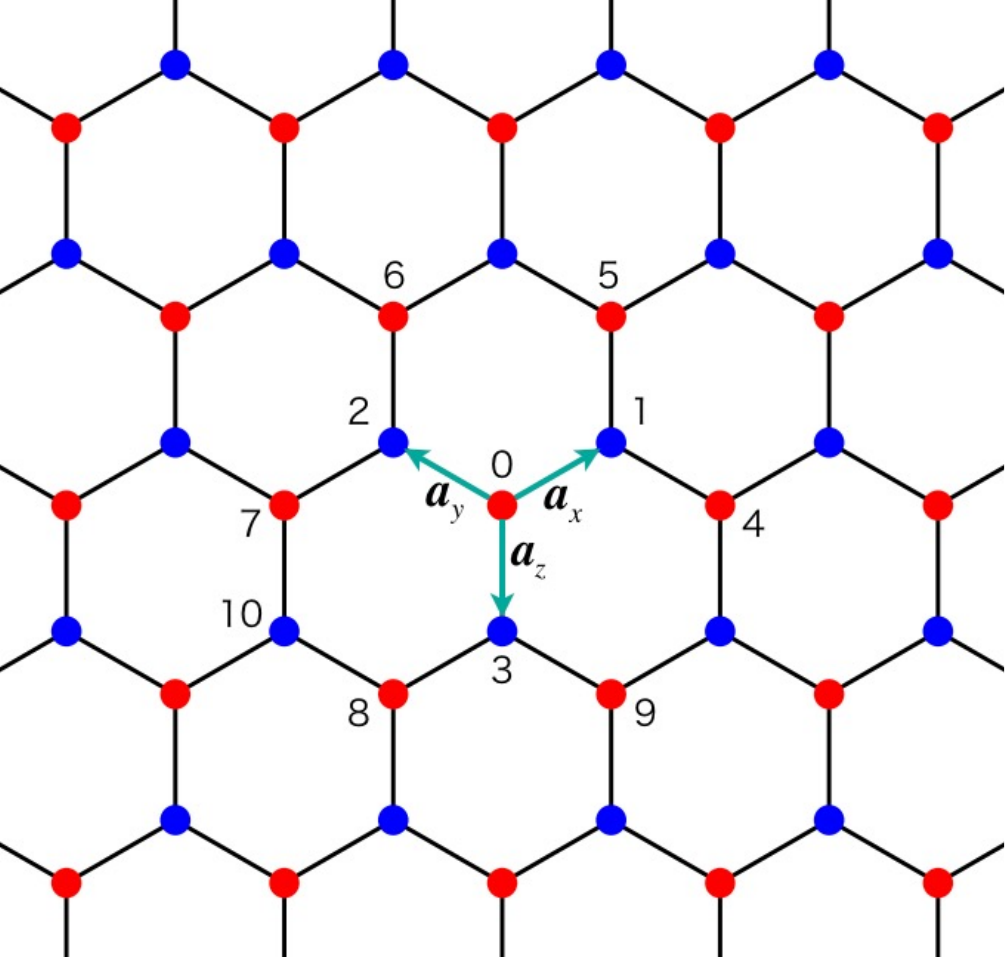}
  \caption{
    \label{fig:honeycomb_lattice}
The honeycomb lattice structure in the Kitaev model, illustrating the nearest-neighbor vectors ${{\bf{a}}_x}$, ${{\bf{a}}_y}$, and ${{\bf{a}}_z}$, corresponding to the $x$, $y$, and $z$ directions, respectively. To facilitate the presentation of the Green's function formalism and the derivation of the equations of motion, several lattice sites are numbered for clarity.
  }
\end{figure}

Now, we apply the formalism described above to the Kitaev model. The Hamiltonian is given by 
    \be
       {\cal H} =  - \sum\limits_{\gamma = x,y,z}  {{J_\gamma }
         \sum\limits_{{{\left\langle {i,j} \right\rangle }_\gamma }} {S_i^\gamma S_j^\gamma } }.
       \ee
In the Kitaev model, the interaction between nearest-neighbor spins is an Ising-like coupling that depends on the direction of the bond. The notation ${\left\langle {i,j} \right\rangle }_\gamma$ refers to nearest-neighbor sites $i$ and $j$ connected by a bond in the $\gamma$ direction. Denoting the coordinate of the $j$-th site as ${\bf r}_j$, the relative position between sites $i$ and $j$ is given by ${{\bf r}_i} - {{\bf r}_j} =  \pm {{\bf{a}}_\gamma }$, where the bond vectors ${{\bf{a}}_\gamma }$ are:
    \bea
        {{\bf{a}}_x} &=& \left( {\frac{1}{2},\frac{1}{{2\sqrt 3 }}} \right)a, \\
        {{\bf{a}}_y} &=& \left( { - \frac{1}{2},\frac{1}{{2\sqrt 3 }}} \right)a, \\
        {{\bf{a}}_z} &=& \left( {0, - \frac{1}{{\sqrt 3 }}} \right)a.
        \eea
Here, $a$ represents the lattice constant. The vectors ${{\bf a}_\gamma }$ are illustrated in Fig.~\ref{fig:honeycomb_lattice}.

For the Green's function, we consider the case involving two spins:
\be
G_{0,j}^x\left( {i{\omega _n}} \right)
\equiv G_{0,j}^{xx}\left( {i{\omega _n}} \right)
= {\left\langle {{S_0^x}}
 \mathrel{\left | {\vphantom {{S_0^x} {S_j^x}}}
 \right. \kern-\nulldelimiterspace}
         {{S_j^x}} \right\rangle _{i{\omega _n}}}.
\label{eq:G0j_x}
\ee
For convenience, we assign numbers to certain lattice sites, as shown in Fig.~\ref{fig:honeycomb_lattice}.

The lattice site $j$ can represent any site and is not restricted to the numbered sites in Fig.~\ref{fig:honeycomb_lattice}. However, the Kitaev model exhibits a special property: the Green's function $G_{0j}^\mu(\tau)$, as defined in Eq.~(\ref{eq:G0j_x}), is non-zero only when $j = 0$ or $j = 1$. Furthermore, we do not need to consider Green's functions of the form
${\left\langle {{S_0^\mu }}
 \mathrel{\left | {\vphantom {{S_0^\mu } {S_j^\nu }}}
 \right. \kern-\nulldelimiterspace}
         {{S_j^\nu }} \right\rangle _{i{\omega _n}}}$
with $\mu \neq \nu$,
as these are identically zero. This property has been rigorously demonstrated at zero temperature \cite{Baskaran2007} and is well-documented in the literature \cite{Motome2020}. We briefly explain this in Appendix~\ref{app:exact-results}.

To derive the equation of motion for the Green's function in Eq.~(\ref{eq:G0j_x}), we first compute the commutator $\left[ S_0^x, {\cal H} \right]$, which is given by
\be
   [S_0^x,{\cal H}] =  - i{J_y}S_0^zS_2^y + i{J_z}S_0^yS_3^z.
   \ee
Since $\left\langle \left[ S_0^x, S_j^x \right] \right\rangle = 0$, Eq.~(\ref{eq:G_EOM1}) simplifies to
\be
i{\omega _n}G_{0,j}^x\left( {i{\omega _n}} \right) =  - i{J_y}{\left\langle {{S_0^zS_2^y}}
  \mathrel{\left | {\vphantom {{S_0^zS_2^y} {S_j^x}}}
    \right. \kern-\nulldelimiterspace}
          {{S_j^x}} \right\rangle _{i{\omega _n}}} + i{J_z}{\left\langle {{S_0^yS_3^z}}
  \mathrel{\left | {\vphantom {{S_0^yS_3^z} {S_j^x}}}
    \right. \kern-\nulldelimiterspace}
          {{S_j^x}} \right\rangle _{i{\omega _n}}}.
\label{eq:Gx_0j_1st}
\ee
The equations of motion for the two Green's functions on the right-hand side are then given by
   \be
   i{\omega _n}{\left\langle {{S_0^zS_2^y}}
 \mathrel{\left | {\vphantom {{S_0^zS_2^y} {S_j^x}}}
 \right. \kern-\nulldelimiterspace}
 {{S_j^x}} \right\rangle _{i{\omega _n}}} = {\left\langle {{\left[ {S_0^zS_2^y,H} \right]}}
 \mathrel{\left | {\vphantom {{\left[ {S_0^zS_2^y,H} \right]} {S_j^x}}}
 \right. \kern-\nulldelimiterspace}
         {{S_j^x}} \right\rangle _{i{\omega _n}}} + \left\langle {\left[ {S_0^zS_2^y,S_j^x} \right]}
   \right\rangle,
   \ee
   \be
   i{\omega _n}{\left\langle {{S_0^yS_3^z}}
 \mathrel{\left | {\vphantom {{S_0^yS_3^z} {S_j^x}}}
 \right. \kern-\nulldelimiterspace}
 {{S_j^x}} \right\rangle _{i{\omega _n}}} = {\left\langle {{\left[ {S_0^yS_3^z,H} \right]}}
 \mathrel{\left | {\vphantom {{\left[ {S_0^yS_3^z,H} \right]} {S_j^x}}}
 \right. \kern-\nulldelimiterspace}
         {{S_j^x}} \right\rangle _{i{\omega _n}}} + \left\langle {\left[ {S_0^yS_3^z,S_j^x} \right]}
   \right\rangle.
   \ee
   
   \begin{widetext}
By computing the commutators that appear on the right-hand side, we obtain
   \bea
   i{\omega _n}{\left\langle {{S_0^zS_2^y}}
 \mathrel{\left | {\vphantom {{S_0^zS_2^y} {S_j^x}}}
 \right. \kern-\nulldelimiterspace}
         {{S_j^x}} \right\rangle _{i{\omega _n}}}
   &=&  - i{J_x}{\left\langle {{S_0^yS_1^xS_2^y}}
 \mathrel{\left | {\vphantom {{S_0^yS_1^xS_2^y} {S_j^x}}}
 \right. \kern-\nulldelimiterspace}
         {{S_j^x}} \right\rangle _{i{\omega _n}}}
   + \frac{1}{4}i{J_y}{\left\langle {S_0^x}
     \middle|
     {{S_j^x}} \right\rangle _{i{\omega _n}}} \nonumber \\
   & &  - i{J_z}{\left\langle {{S_0^zS_2^xS_6^z}}
 \mathrel{\left | {\vphantom {{S_0^zS_2^xS_6^z} {S_j^x}}}
 \right. \kern-\nulldelimiterspace}
 {{S_j^x}} \right\rangle _{i{\omega _n}}} + i{J_x}{\left\langle {{S_0^zS_2^zS_7^x}}
 \mathrel{\left | {\vphantom {{S_0^zS_2^zS_7^x} {S_j^x}}}
 \right. \kern-\nulldelimiterspace}
         {{S_j^x}} \right\rangle _{i{\omega _n}}} \nonumber \\
   & &  - i{\delta _{2,j}}\left\langle {S_0^zS_2^z} \right\rangle  + i{\delta _{0,j}}\left\langle {S_0^yS_2^y} \right\rangle,
   \label{eq:G_EOM_2nd_1}
   \eea
   and
   \bea
   i{\omega _n}{\left\langle {{S_0^yS_3^z}}
 \mathrel{\left | {\vphantom {{S_0^yS_3^z} {S_j^x}}}
 \right. \kern-\nulldelimiterspace}
         {{S_j^x}} \right\rangle _{i{\omega _n}}}
   &=& i{J_x}{\left\langle {{S_0^zS_1^xS_3^z}}
 \mathrel{\left | {\vphantom {{S_0^zS_1^xS_3^z} {S_j^x}}}
 \right. \kern-\nulldelimiterspace}
         {{S_j^x}} \right\rangle _{i{\omega _n}}}
   - \frac{1}{4}i{J_z}{\left\langle {{S_0^x}}
     \middle|
     {{S_j^x}} \right\rangle _{i{\omega _n}}} \nonumber \\
   & & - i{J_x}{\left\langle {{S_0^yS_3^yS_8^x}}
 \mathrel{\left | {\vphantom {{S_0^yS_3^yS_8^x} {S_j^x}}}
 \right. \kern-\nulldelimiterspace}
 {{S_j^x}} \right\rangle _{i{\omega _n}}} + i{J_y}{\left\langle {{S_0^yS_3^xS_9^y}}
 \mathrel{\left | {\vphantom {{S_0^yS_3^xS_9^y} {S_j^x}}}
 \right. \kern-\nulldelimiterspace}
         {{S_j^x}} \right\rangle _{i{\omega _n}}} \nonumber \\
   & &  + i{\delta _{3,j}}\left\langle {S_0^yS_3^y} \right\rangle  - i{\delta _{0,j}}\left\langle {S_0^zS_3^z} \right\rangle.
   \label{eq:G_EOM_2nd_2}   
   \eea

Next, we apply the decoupling approximation. There are several ways to decouple the Green's functions involving four spins that appear in Eqs.~(\ref{eq:G_EOM_2nd_1}) and (\ref{eq:G_EOM_2nd_2}), and to introduce parameters that enhance the accuracy of the approximation. However, the specific decoupling scheme is uniquely determined based on physical considerations, particularly since $j$ is limited to either $j=0$ or $j=1$.

As an example, let us examine the four-spin Green's function in the first term of Eq.~(\ref{eq:G_EOM_2nd_1}):
\be
{\left\langle {{S_0^yS_1^xS_2^y}}
 \mathrel{\left | {\vphantom {{S_0^yS_1^xS_2^y} {S_j^x}}}
 \right. \kern-\nulldelimiterspace}
 {{S_j^x}} \right\rangle _{i{\omega _n}}} = {\delta _{j,0}}{\left\langle {{S_0^yS_1^xS_2^y}}
 \mathrel{\left | {\vphantom {{S_0^yS_1^xS_2^y} {S_0^x}}}
 \right. \kern-\nulldelimiterspace}
 {{S_0^x}} \right\rangle _{i{\omega _n}}} + {\delta _{j,1}}{\left\langle {{S_0^yS_1^xS_2^y}}
 \mathrel{\left | {\vphantom {{S_0^yS_1^xS_2^y} {S_1^x}}}
 \right. \kern-\nulldelimiterspace}
         {{S_1^x}} \right\rangle _{i{\omega _n}}}.
\label{eq:SSS_S}
\ee
We extract two spins from the product $S_0^y S_1^x S_2^y$ such that the correlation function of those spins and the remaining Green's function are non-zero. This procedure is uniquely determined by the considerations discussed in Appendix \ref{app:exact-results}. However, care must be taken when introducing parameters during the decoupling process.

The decoupled form of Eq.~(\ref{eq:SSS_S}) is given by
\be
{\left\langle {{S_0^yS_1^xS_2^y}}
 \mathrel{\left | {\vphantom {{S_0^yS_1^xS_2^y} {S_j^x}}}
 \right. \kern-\nulldelimiterspace}
         {{S_j^x}} \right\rangle _{i{\omega _n}}}
\simeq 
{\delta _{j,0}} \eta \left\langle {S_0^yS_2^y} \right\rangle {\left\langle {{S_1^x}}
 \mathrel{\left | {\vphantom {{S_1^x} {S_0^x}}}
 \right. \kern-\nulldelimiterspace}
         {{S_0^x}} \right\rangle _{i{\omega _n}}}
+ {\delta _{j,1}}\alpha \left\langle {S_0^yS_2^y} \right\rangle {\left\langle {{S_1^x}}
 \mathrel{\left | {\vphantom {{S_1^x} {S_1^x}}}
 \right. \kern-\nulldelimiterspace}
         {{S_1^x}} \right\rangle _{i{\omega _n}}}.
\label{eq:decoupling_alpha_eta}
\ee
Note that we introduce different parameters, $\eta$ and $\alpha$.
  The need to introduce different parameters is not immediately apparent from the form of
  ${\left\langle {{S_0^yS_1^xS_2^y}}
 \mathrel{\left | {\vphantom {{S_0^yS_1^xS_2^y} {S_j^x}}}
 \right. \kern-\nulldelimiterspace}
         {{S_j^x}} \right\rangle _{i{\omega _n}}}$.
  To clarify this, we consider the corresponding Green's function in imaginary time:
  \be
  \begin{split}
    \left\langle {S_0^y\left( \tau  \right)S_1^x\left( \tau  \right)S_2^y\left( \tau  \right)S_j^x\left( 0 \right)} \right\rangle
    &=  - {\left\langle {{S_0^yS_1^xS_2^y}}
 \mathrel{\left | {\vphantom {{S_0^yS_1^xS_2^y} {S_j^x}}}
 \right. \kern-\nulldelimiterspace}
         {{S_j^x}} \right\rangle _\tau } \nonumber \\
    &= - \frac{1}{\beta }\sum\limits_{i{\omega _n}} {{e^{ - i{\omega _n}\tau }}{{\left\langle {{S_0^yS_1^xS_2^y}}
 \mathrel{\left | {\vphantom {{S_0^yS_1^xS_2^y} {S_j^x}}}
 \right. \kern-\nulldelimiterspace}
 {{S_j^x}} \right\rangle }_{i{\omega _n}}}}.
    \end{split}
  \ee
  This requires us to determine how to decouple
  $\left\langle {S_0^y\left( \tau  \right)S_1^x\left( \tau  \right)S_2^y\left( \tau  \right)S_j^x\left( 0 \right)} \right\rangle$.
  Since this quantity is required at $\tau = +0$ to solve the self-consistent equations, it suffices to focus on the decoupling of the following correlation function:
  \be
  \left\langle {S_0^yS_1^xS_2^yS_j^x} \right\rangle.
  \ee
  This correlation function is non-zero only for $j=0,1$.
  For the case of $j=0$, we find
  \be
  \left\langle {S_0^yS_1^xS_2^yS_0^x} \right\rangle  =  - \frac{i}{2}\left\langle {S_0^zS_1^xS_2^y} \right\rangle.
  \ee
  For the case of $j=1$, we find
  \be
  \left\langle {S_0^yS_1^xS_2^yS_1^x} \right\rangle  = \frac{1}{4}\left\langle {S_0^yS_2^y} \right\rangle.
  \ee
  Clearly, these correlation functions take different values. To account for this difference, it is necessary to introduce distinct parameters in the decoupling procedure.

For the other two four-spin Green's functions on the right-hand side of Eq.~(\ref{eq:G_EOM_2nd_1}), the decoupling is uniquely determined by the fact that $G_{i,j}^{xz}\left( i{\omega _n} \right) = 0$. Therefore, we obtain
\be
  {\left\langle {{S_0^zS_2^xS_6^z}}
 \mathrel{\left | {\vphantom {{S_0^zS_2^xS_6^z} {S_j^x}}}
 \right. \kern-\nulldelimiterspace}
 {{S_j^x}} \right\rangle _{i{\omega _n}}} \simeq \left\langle {S_0^zS_6^z} \right\rangle {\left\langle {{S_2^x}}
 \mathrel{\left | {\vphantom {{S_2^x} {S_j^x}}}
 \right. \kern-\nulldelimiterspace}
         {{S_j^x}} \right\rangle _{i{\omega _n}}} = 0,
  \ee
  \be
    {\left\langle {{S_0^zS_2^zS_7^x}}
 \mathrel{\left | {\vphantom {{S_0^zS_2^zS_7^x} {S_j^x}}}
 \right. \kern-\nulldelimiterspace}
 {{S_j^x}} \right\rangle _{i{\omega _n}}} \simeq \left\langle {S_0^zS_2^z} \right\rangle {\left\langle {{S_7^x}}
 \mathrel{\left | {\vphantom {{S_7^x} {S_j^x}}}
 \right. \kern-\nulldelimiterspace}
         {{S_j^x}} \right\rangle _{i{\omega _n}}} = 0.
    \ee
      Note that
      $\left\langle S_0^z S_6^z \right\rangle = 0$ and $\left\langle S_0^z S_2^z \right\rangle = 0$.
      Thus, after applying the decoupling procedure, we obtain
      \be
      i{\omega _n}{\left\langle {{S_0^zS_2^y}}
 \mathrel{\left | {\vphantom {{S_0^zS_2^y} {S_j^x}}}
 \right. \kern-\nulldelimiterspace}
         {{S_j^x}} \right\rangle _{i{\omega _n}}} =  - i{J_x}{c_y}\left[ {{\delta _{j,0}}\eta G_{1,0}^x\left( {i{\omega _n}} \right) + {\delta _{j,1}}\alpha G_{1,1}^x\left( {i{\omega _n}} \right)} \right] + \frac{1}{4}i{J_y}G_{0,j}^x\left( {i{\omega _n}} \right) + i{\delta _{0,j}}{c_y},
      \ee
where we define the correlation function      
      \be
      {c_y} = \left\langle {S_0^yS_2^y} \right\rangle.
      \ee
We also define the following correlation functions:
      \bea
          {c_x} &=& \left\langle {S_0^xS_1^x} \right\rangle,\\
          {c_z} &=& \left\langle {S_0^zS_3^z} \right\rangle.
          \eea

Applying similar decoupling procedures to Eq.~(\ref{eq:G_EOM_2nd_2}), we obtain          
\be
i{\omega _n}{\left\langle {{S_0^yS_3^z}}
 \mathrel{\left | {\vphantom {{S_0^yS_3^z} {S_j^x}}}
 \right. \kern-\nulldelimiterspace}
         {{S_j^x}} \right\rangle _{i{\omega _n}}} = i{J_x}{c_z}\left[ {{\delta _{j,0}}\eta G_{1,0}^x\left( {i{\omega _n}} \right) + {\delta _{j,1}}\alpha G_{1,1}^x\left( {i{\omega _n}} \right)} \right] - \frac{1}{4}i{J_z}G_{0,j}^x\left( {i{\omega _n}} \right) - i{\delta _{0,j}}{c_z}.
\ee
By combining Eqs.~(\ref{eq:Gx_0j_1st}), (\ref{eq:G_EOM_2nd_1}), and (\ref{eq:G_EOM_2nd_2}), we derive
\be
\left[ {{{\left( {i{\omega _n}} \right)}^2} - \frac{1}{4}\left( {J_y^2 + J_z^2} \right)} \right]G_{0,j}^x\left( {i{\omega _n}} \right) = \left( {{J_y}{c_y} + {J_z}{c_z}} \right)
\left[ { - {\delta _{j,0}}\eta {J_x}G_{1,0}^x\left( {i{\omega _n}} \right) - {\delta _{j,1}}\alpha {J_x}G_{1,1}^x\left( {i{\omega _n}} \right) + {\delta _{0,j}}} \right].
\label{eq:Gx_0j_sol}
\ee
\\
Since this equation involves the Green's functions $G_{1,j}^x\left( i{\omega _n} \right)$ on the right-hand side, we must consider the equation of motion for $G_{1,j}^x\left( i{\omega _n} \right)$. The calculation proceeds similarly to that of $G_{0,j}^x\left( i{\omega _n} \right)$. After applying the decoupling approximation, we obtain
\be
\left[ {{{\left( {i{\omega _n}} \right)}^2} - \frac{1}{4}\left( {J_y^2 + J_z^2} \right)} \right]G_{1,j}^x\left( {i{\omega _n}} \right) = \left( {{J_y}{c_y} + {J_z}{c_z}} \right)\left[ { - {\delta _{j,0}}{J_x}\alpha G_{0,0}^x\left( {i{\omega _n}} \right) - {\delta _{j,1}}{J_x}\eta G_{0,1}^x\left( {i{\omega _n}} \right) + {\delta _{1,j}}} \right]
\label{eq:Gx_1j_sol}
\ee
\\
Setting $j=0$ and $j=1$ in Eqs.~(\ref{eq:Gx_0j_sol}) and (\ref{eq:Gx_1j_sol}), we derive
\bea
\left[ {{{\left( {i{\omega _n}} \right)}^2} - \frac{1}{4}\left( {J_y^2 + J_z^2} \right)} \right]G_{0,0}^x\left( {i{\omega _n}} \right)
&=& \left( {{J_y}{c_y} + {J_z}{c_z}} \right)\left[ { - \eta {J_x}G_{1,0}^x\left( {i{\omega _n}} \right) + 1} \right], \\
\left[ {{{\left( {i{\omega _n}} \right)}^2} - \frac{1}{4}\left( {J_y^2 + J_z^2} \right)} \right]G_{0,1}^x\left( {i{\omega _n}} \right)
&=&  - \left( {{J_y}{c_y} + {J_z}{c_z}} \right)\alpha {J_x}G_{1,1}^x\left( {i{\omega _n}} \right), \\
\left[ {{{\left( {i{\omega _n}} \right)}^2} - \frac{1}{4}\left( {J_y^2 + J_z^2} \right)} \right]G_{1,0}^x\left( {i{\omega _n}} \right)
&=&  - \left( {{J_y}{c_y} + {J_z}{c_z}} \right){J_x}\alpha G_{0,0}^x\left( {i{\omega _n}} \right),
\\
\left[ {{{\left( {i{\omega _n}} \right)}^2} - \frac{1}{4}\left( {J_y^2 + J_z^2} \right)} \right]G_{1,1}^x\left( {i{\omega _n}} \right)
&=& \left( {{J_y}{c_y} + {J_z}{c_z}} \right)\left[ { - {J_x}\eta G_{0,1}^x\left( {i{\omega _n}} \right) + 1} \right].
\eea
Using the fact that $G_{0,0}^x\left( i{\omega _n} \right) = G_{1,1}^x\left( i{\omega _n} \right)$, the solution to these equations is
\bea
G_{0,0}^x\left( {i{\omega _n}} \right)
&=& G_{1,1}^x\left( {i{\omega _n}} \right) = \frac{{\left( {{J_y}{c_y} + {J_z}{c_z}} \right)\left[ {{{\left( {i{\omega _n}} \right)}^2} - \frac{1}{4}\left( {J_y^2 + J_z^2} \right)} \right]}}{{{{\left[ {{{\left( {i{\omega _n}} \right)}^2} - \frac{1}{4}\left( {J_y^2 + J_z^2} \right)} \right]}^2} - \eta \alpha J_x^2{{\left( {{J_y}{c_y} + {J_z}{c_z}} \right)}^2}}},
\label{eq:sol_G00_G11_iwn}
\\
G_{1,0}^x\left( {i{\omega _n}} \right)
&=& G_{0,1}^x\left( {i{\omega _n}} \right) =  - \frac{{\alpha {J_x}{{\left( {{J_y}{c_y} + {J_z}{c_z}} \right)}^2}}}{{{{\left[ {{{\left( {i{\omega _n}} \right)}^2} - \frac{1}{4}\left( {J_y^2 + J_z^2} \right)} \right]}^2} - \eta \alpha J_x^2{{\left( {{J_y}{c_y} + {J_z}{c_z}} \right)}^2}}}.
\label{eq:sol_G10_G01_iwn}
\eea
\end{widetext}
We find analogous expressions for $G_{i,j}^y\left( i{\omega _n} \right)$ with $i,j = 0,2$ and $G_{i,j}^z\left( i{\omega _n} \right)$ with $i,j = 0,3$.

From the poles of the Green's functions, we identify the excitation energies as
\bea
E_ \pm ^x &=& \sqrt {\frac{1}{4}\left( {J_y^2 + J_z^2} \right) \pm \sqrt {\eta \alpha } \left| {{J_x}} \right|\left| {{J_y}{c_y} + {J_z}{c_z}} \right|},\\
E_ \pm ^y &=& \sqrt {\frac{1}{4}\left( {J_z^2 + J_x^2} \right) \pm \sqrt {\eta \alpha } \left| {{J_y}} \right|\left| {{J_z}{c_z} + {J_x}{c_x}} \right|},\\
E_ \pm ^z &=& \sqrt {\frac{1}{4}\left( {J_x^2 + J_y^2} \right) \pm \sqrt {\eta \alpha } \left| {{J_z}} \right|\left| {{J_x}{c_x} + {J_y}{c_y}} \right|}.
\eea

Now, we consider the case where $J_z \neq J_x = J_y$. From Eqs.~(\ref{eq:sol_G00_G11_iwn}) and (\ref{eq:sol_G10_G01_iwn}), we observe that the Green's functions are determined by the correlation functions $c_z$, $c_x = c_y$, and the decoupling parameters $\alpha$ and $\eta$. These parameters are derived from the Green's functions through
\be
\left\langle {S_0^\gamma S_j^\gamma } \right\rangle
=  - G_{0,j}^\gamma \left( {\tau  =  + 0} \right).
\label{eq:corr_G_relation}
\ee
Setting $j=0,3$ for $\gamma = z$ and $j=0,1$ for $\gamma = x$, we obtain
\bea
\frac{1}{4} &=& \left\langle {S_0^zS_0^z} \right\rangle
=  - G_{0,0}^z\left( {\tau  =  + 0} \right),
\label{eq:sceq_Gz00}
\\
{c_z} &=& \left\langle {S_0^zS_3^z} \right\rangle
=  - G_{0,3}^z\left( {\tau  =  + 0} \right),
\label{eq:sceq_Gz01}
\\
\frac{1}{4} &=& \left\langle {S_0^xS_0^x} \right\rangle
=  - G_{0,0}^x\left( {\tau  =  + 0} \right),
\label{eq:sceq_Gx00}
\\
{c_x} &=& \left\langle {S_0^xS_1^x} \right\rangle
=  - G_{0,1}^x\left( {\tau  =  + 0} \right).
\label{eq:sceq_Gx01}
\eea
These equations represent the self-consistent conditions.
\begin{widetext}   
The explicit forms of the right-hand sides are given by
  \be
  G_{0,0}^x\left( {\tau  =  + 0} \right)
  = \frac{{{J_y}{c_y} + {J_z}{c_z}}}{2}\left[ {F\left( {E_ + ^x} \right) + F\left( {E_ - ^x} \right)} \right],
  \label{eq:G_00x_tau0}
  \ee
  \be
  G_{0,0}^z\left( {\tau  =  + 0} \right)
  = \frac{{{J_x}{c_x} + {J_y}{c_y}}}{2}\left[ {F\left( {E_ + ^z} \right) + F\left( {E_ - ^z} \right)} \right],
  \ee
  \be
  G_{0,1}^x\left( {\tau  =  + 0} \right)
  =  - \frac{{\left| {{J_x}} \right|\left| {{J_y}{c_y} + {J_z}{c_z}} \right|}}{2}\sqrt {\frac{\alpha }{\eta }} \left[ {F\left( {E_ + ^x} \right) - F\left( {E_ - ^x} \right)} \right],
  \ee
  \be
  G_{0,3}^z\left( {\tau  =  + 0} \right)
  =  - \frac{{\left| {{J_z}} \right|\left| {{J_x}{c_x} + {J_y}{c_y}} \right|}}{2}\sqrt {\frac{\alpha }{\eta }} \left[ {F\left( {E_ + ^z} \right) - F\left( {E_ - ^z} \right)} \right].
  \label{eq:G_03z_tau0}
  \ee
  Here, the function $F(E)$ is defined as  
  \be
  F\left( E \right) =  - \frac{1}{{2E\tanh \left( {\frac{{\beta E}}{2}} \right)}}.
  \ee
\end{widetext}

  Note that the Matsubara summation,
  which is required to derive Eqs.~(\ref{eq:G_00x_tau0}) to (\ref{eq:G_03z_tau0}),
  is performed analytically.
  As an example, Eq.~(\ref{eq:sol_G00_G11_iwn}) can be rewritten as
  \be
  \begin{split}
  G_{0,0}^x\left( {i{\omega _n}} \right)
  &= \frac{{{J_y}{c_y} + {J_z}{c_z}}}{2}
  \left[ \frac{1}{{{{\left( {i{\omega _n}} \right)}^2}
        - {{\left( {E_ + ^x} \right)}^2}}}
    \right. \\
    & \left. + \frac{1}{{{{\left( {i{\omega _n}} \right)}^2}
          - {{\left( {E_ - ^x} \right)}^2}}}
    \right].
  \end{split}
  \label{eq:G_00x_rewritten}
  \ee
  To evaluate this expression, we use the following formula
  for the bosonic Matsubara summation:  
  \begin{align}
    & \frac{1}{\beta }\sum\limits_{i{\omega _n}}
    {{e^{ - i{\omega _n}\tau }}} \frac{1}{{{{\left( {i{\omega _n}} \right)}^2} - {E^2}}}
    \nonumber \\
    = & \frac{1}{{2\beta E}}
    \sum\limits_{i{\omega _n}} {{e^{ - i{\omega _n}\tau }}}
    \left( {\frac{1}{{i{\omega _n} - E}} - \frac{1}{{i{\omega _n} + E}}} \right)
    \nonumber \\
    = & - \frac{{{e^{ - \tau E}}\left[ {n\left( E \right) + 1} \right] - {e^{\tau E}}\left[ {n\left( { - E} \right) + 1} \right]}}{{2E}},
  \end{align}
    where $n\left( E \right) = 1/\left( {{e^{\beta E}} - 1} \right)$ is
    the Bose distribution function.
    Applying this formula to Eq.~(\ref{eq:G_00x_rewritten}), we obtain
    Eq.~(\ref{eq:G_00x_tau0}) at $\tau = +0$.

  In the isotropic case, where $J_x = J_y = J_z$, there are two self-consistent equations, Eqs.~(\ref{eq:sceq_Gz00}) and (\ref{eq:sceq_Gz01}). To ensure that the number of parameters matches the number of self-consistent equations, we set $\alpha = \eta$. However, this is not an ideal approximation, as we will demonstrate that the error is larger for the isotropic case compared to the anisotropic cases. A more accurate calculation could be achieved by treating $\alpha$ and $\eta$ as independent parameters, allowing one to be adjusted to improve the accuracy of the results. Nonetheless, we do not pursue this refinement in the present study.


In general, solving a system of non-linear equations with several parameters is a non-trivial task. However, one can utilize the high-temperature expansion results to provide initial values for the parameters $c_x$, $c_z$, $\alpha$, and $\eta$. The lowest-order result of the high-temperature expansion is
\be
c_\gamma ^{{\rm{HTE}}} = \frac{1}{{16}}\beta {J_\gamma }.
\ee
  For the initial value of $\alpha$, we set $\alpha = 1$,
  which is derived from the approximate forms of Eqs.~(\ref{eq:sceq_Gz00}) to (\ref{eq:sceq_Gx01}),
  neglecting terms of the order of $1/T^2$ at high temperatures.
  Additionally, we find that the value of $\eta$ is irrelevant,
  based on these approximate forms, when solving the self-consistent equations at high temperatures.
Starting from a high temperature, say $T \sim 10 J_x$ or $10 J_z$, we solve the self-consistent equations and then gradually lower the temperature, using the parameters obtained from the previous step as initial values for the next.

In the following analysis, we focus on the case where $J_x/J = J_y/J = \xi$ and $J_z/J = 3 - 2\xi$
with $J$ as the unit of energy.

\subsection{Computation of physical quantities}
We now demonstrate how physical quantities are computed from the Green's functions. The correlation functions are directly obtained from the Green's functions using Eq.~(\ref{eq:corr_G_relation}). As noted earlier, for the Kitaev model with $J_x = J_y$, the non-vanishing correlation functions are limited to $c_z$ and $c_x = c_y$.

Since the interactions in the Kitaev model are restricted to nearest-neighbor sites, the internal energy, $E$, can be readily computed as
\be
E = \left\langle {\cal H} \right\rangle
=  - \frac{N}{2}\left( {{J_x}{c_x} + {J_y}{c_y} + {J_z}{c_z}} \right).
\ee
Here, $N$ represents the number of lattice sites. For the case where $J_x = J_y$, this simplifies to
\be
E =  - \frac{N}{2}\left( {2{J_x}{c_x} + {J_z}{c_z}} \right).
\ee
Naively, the specific heat can be computed from the temperature derivative of $E$, as demonstrated in Ref.~\onlinecite{Takegami2024}. However, special care is required for an accurate calculation of the specific heat. In particular, four-spin correlation functions are necessary for improved precision. This more detailed calculation is left for future research.

The spin susceptibility can also be directly computed from the Green's function\cite{Mahan1990}:
\be
   {\chi ^{\mu \nu }} = \frac{1}{N}\sum\limits_{i,j} {\int_0^\beta  {d\tau } \left\langle {S_i^\mu \left( \tau  \right)S_j^\nu \left( 0 \right)} \right\rangle }.
   \ee
As shown in Appendix \ref{app:exact-results}, the off-diagonal elements ${\chi ^{\mu \nu }} = 0$ for $\mu \neq \nu$. However, the diagonal components, ${\chi ^{zz}}$ and ${\chi ^{xx} = {\chi ^{yy}}}$, are non-zero. In terms of the Green's functions, we have
   \be
      {\chi ^{\gamma \gamma }} =  - \sum\limits_j {G_{0,j}^\gamma \left( {i{\omega _n} = 0} \right)}.
      \ee
Therefore, we arrive at
      \be
         {\chi ^{zz}} =  - G_{0,0}^z\left( {i{\omega _n} = 0} \right) - G_{0,3}^z\left( {i{\omega _n} = 0} \right),
         \ee
         \be
            {\chi ^{xx}} =  - G_{0,0}^x\left( {i{\omega _n} = 0} \right) - G_{0,1}^x\left( {i{\omega _n} = 0} \right).
            \ee
For the isotropic case where $J_x = J_y = J_z = J$, this simplifies further to            
            \be
               {\chi ^{zz}} = {\chi ^{xx}} = \frac{{4{c_z}}}{{J\left( {1 - 4\alpha {c_z}} \right)}}.
               \ee
               It can be shown that this formula agrees with the Curie-Weiss law at high temperatures by setting $\alpha = 1$ and $c_z = J/(16T)$.
               
The dynamical spin structure factor is defined as
\bea
    {S^{\mu \nu }}\left( {{\bf{q}},\omega } \right)
    &=& \frac{1}{N}\sum\limits_{i,j} {\int_{ - \infty }^\infty  {\frac{{dt}}{{2\pi }}} } {e^{i\left[ {\omega t - {\bf{q}} \cdot \left( {{{\bf{r}}_i} - {{\bf{r}}_j}} \right){{\bf{R}}_{ij}}} \right]}}
    \nonumber \\
    & & \times
    \left\langle {S_i^\mu \left( t \right)S_j^\nu \left( 0 \right)} \right\rangle.
    \eea
In terms of the Green's function, this expression becomes    
   \bea
       {S^{\mu \nu }}\left( {{\bf{q}},\omega } \right)
       &=&  - \frac{1}{\pi }\frac{1}{{1 - {e^{ - \beta \omega }}}}
       \frac{1}{N}\sum\limits_{i,j}
      {{e^{ - i{\bf{q}} \cdot {{\bf{R}}_{ij}}}}}
      \nonumber \\
      & & \times {\mathop{\rm Im}\nolimits} G_{ij}^{\mu \nu }
      \left( {i{\omega _n} \to \omega  + i\delta } \right).
      \label{eq:Sqw-mu-mu}
      \eea
        Hereafter, we set $\hbar = 1$, and $\delta$ will denote an infinitesimal positive number.
      Since the spin structure factor satisfies
      ${S^{\mu \nu }}\left( {{\bf{q}},\omega } \right) = \delta_{\mu,\nu}
      {S^{\mu \mu }}\left( {{\bf{q}},\omega } \right)$
      and $i$ and $j$ are restricted to nearest-neighbor sites on the $\mu$ bond,
      we can simplify the expression as
      \bea
          {S^{\mu \mu }}\left( {{\bf{q}},\omega } \right)
          &=&  - \frac{1}{\pi }\frac{1}{{1 - {e^{ - \beta \omega }}}}\left[ {{\mathop{\rm Im}\nolimits} G_{0,0}^\mu \left( {i{\omega _n} \to \omega  + i\delta } \right)} \right.
            \nonumber \\
            & &
            \hspace{-2em}
            \left. { + \cos \left( {{\bf{q}} \cdot {{\bf{a}}_\mu }} \right){\mathop{\rm Im}\nolimits} G_{0,{i_\mu }}^\mu \left( {i{\omega _n} \to \omega  + i\delta } \right)} \right],
      \label{eq:Sqw_each}
      \eea
      where $i_x = 1$, $i_y = 2$, and $i_z = 3$.
      The total dynamical structure factor is then given by      
      \be
      S\left( {{\bf{q}},\omega } \right) = \sum\limits_{\mu  = x,y,z} {{S^{\mu \mu }}\left( {{\bf{q}},\omega } \right)}.
      \label{eq:Sqw}      
      \ee
      In the numerical calculations presented below, we assign a small positive value to $\delta$ to account for impurity effects phenomenologically.
      Note that the minus sign on the right-hand side of Eq.~(\ref{eq:Sqw-mu-mu})
      arises from the definition of the Green's function in Eq.~(\ref{eq:Gij}).

\section{Correlation functions at zero temperature}
\label{sec:corr-func-T0}
\begin{figure}[htbp]
  \includegraphics[width=0.9 \linewidth, angle=0]{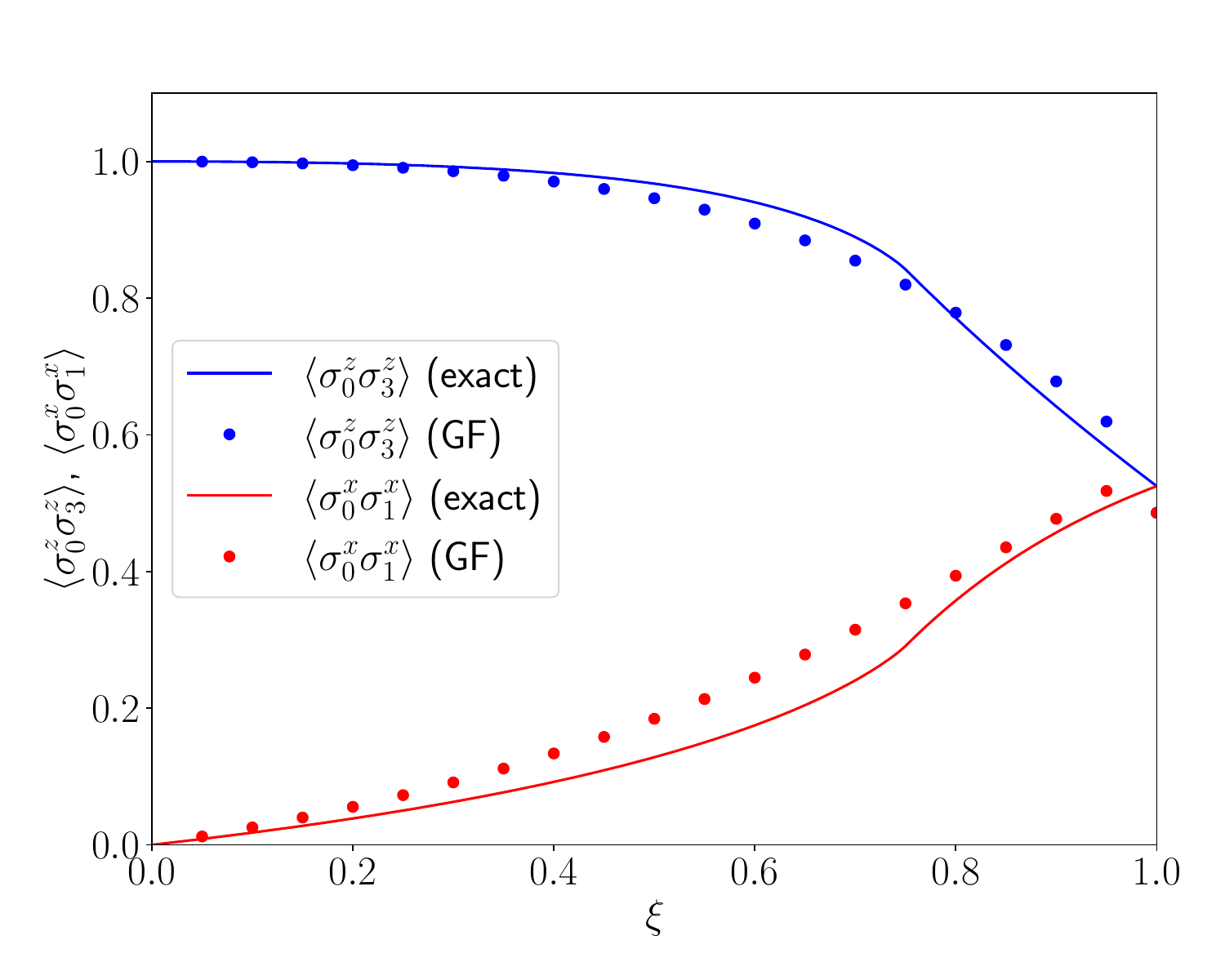}
  \caption{
    \label{fig:cz_cx_T0}
    $\xi$ dependence of the nearest-neighbor correlation functions $c_z$ and $c_x$ at $T=0$, compared with the exact results obtained from the Majorana fermion method \cite{Baskaran2007}. The Green's function results are calculated by initially solving the self-consistent equation at high temperatures, using values from the high-temperature expansion as the starting point, and then extrapolating to zero temperature.
  }
\end{figure}

\begin{figure}[htbp]
  \includegraphics[width=0.9 \linewidth, angle=0]{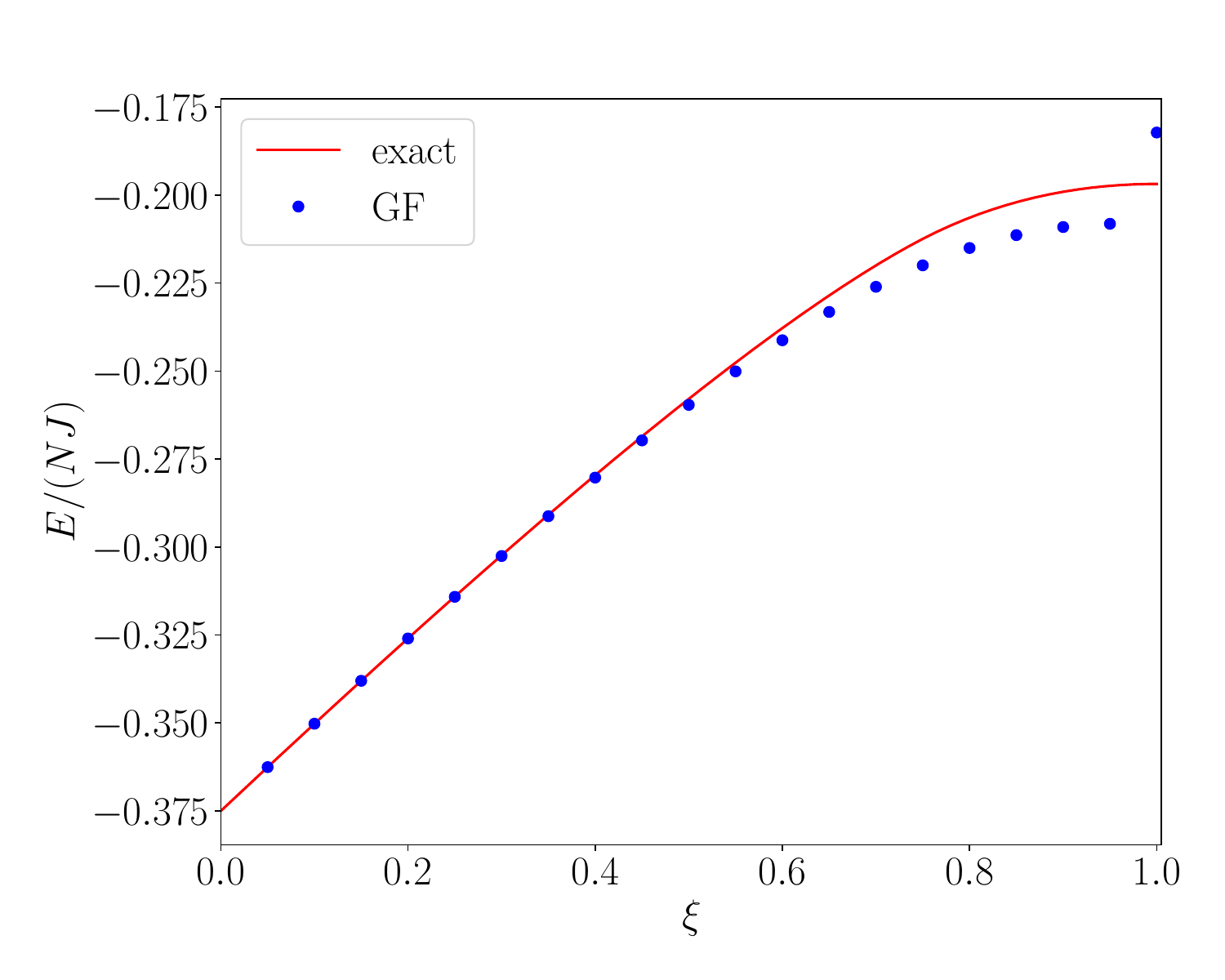}
  \caption{
    \label{fig:E-T0}
    $\xi$ dependence of the ground-state energy. The solid line represents the exact values calculated from the correlation functions obtained via the Majorana fermion method. The points represent the results from the Green's function approach, which show good agreement with the exact values, particularly for $\xi < 0.6$. Note that the value at $\xi = 1$ is obtained under the assumption $\eta = \alpha$.
  }
\end{figure}

  In general, the spin Green's function approach is highly accurate at high temperatures but becomes less precise at low temperatures. To assess the accuracy of this method for the Kitaev model, we focus on the correlation functions at zero temperature, where the largest deviations from exact results are expected. Figure \ref{fig:cz_cx_T0} shows the $\xi$ dependence of the nearest-neighbor correlation functions, $c_z$ and $c_x$, at $T=0$, compared with the exact results obtained via the Majorana fermion method \cite{Baskaran2007}. In Fig.~\ref{fig:E-T0}, we plot the energy as a function of $\xi$, comparing it with the exact values computed from the correlation functions derived using the Majorana fermion representation.

  For the Green's function calculations, we begin at relatively high temperatures, solving the self-consistent equation accurately using initial values from the high-temperature expansion. The temperature is then decreased step by step, and the zero-temperature values are determined through extrapolation.

  The Green's function method yields results that closely match the exact values across the entire range of $\xi$ for the correlation functions. The method is particularly accurate for ground-state energy calculations when $\xi < 0.6$. In this parameter region, the accuracy of the Green's function method increases as the temperature rises, leading to reliable results.

For the isotropic case, the relative error in the energy calculation is 7.4\%,
primarily due to the approximation $\eta = \alpha$, a constraint imposed by the self-consistent calculation rather than by the inherent correlation functions. In this study, we applied the decoupling approximation at the second-order equation of motion. It is possible that applying decoupling at higher orders of the equation of motion would further improve the accuracy. This remains a subject for future investigation.

\section{Finite-Temperature Analysis}
\label{sec:finiteT}
Now, we present the results at finite temperatures. As the temperature increases, we expect more accurate results compared to those at zero temperature.

\subsection{Internal energy}
\begin{figure}[htbp]
  \includegraphics[width=0.9 \linewidth, angle=0]{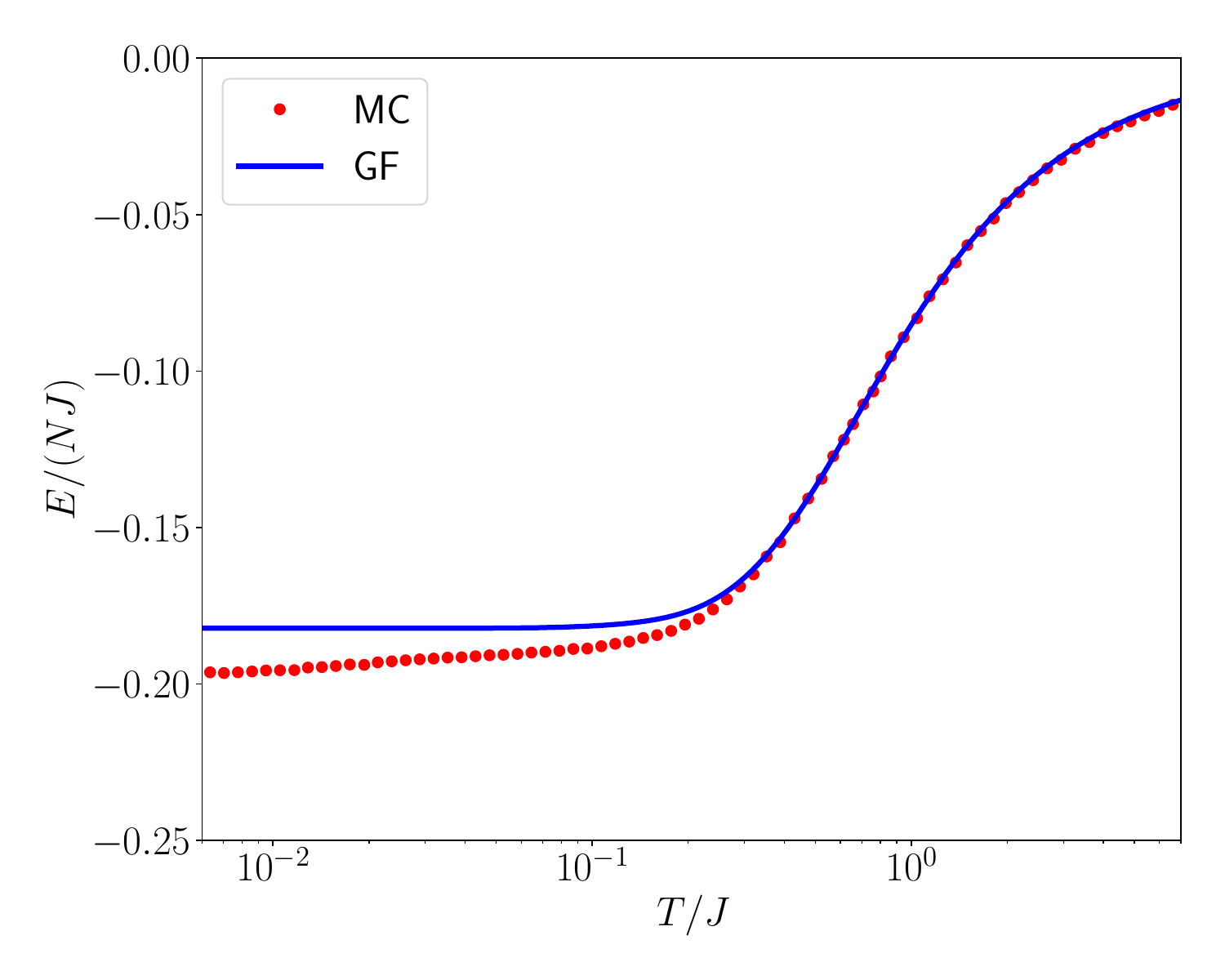}
  \caption{
    \label{fig:FM-iso-MotomeNasu2020Fig15a}
    Temperature dependence of the internal energy for the isotropic case.
    The results are compared with a Majorana-based numerical simulations\cite{Motome2020}. The comparison shows excellent agreement for $T/J > 0.3$, while a discrepancy of approximately 10\% is observed at zero temperature.         }
\end{figure}

In Fig.~\ref{fig:FM-iso-MotomeNasu2020Fig15a}, we show the temperature dependence of the internal energy for the isotropic case, $J_x = J_y = J_z \equiv J$. The results are compared with
a Majorana-based numerical simulation\cite{Nasu2015, Motome2020}.
At zero temperature,
the Majorana-based numerical simulation is
consistent with those from the numerical diagonalization of a finite system \cite{Kitaev2006}. 

The Green's function approach yields results that are in good agreement
with the Majorana-based numerical simulation for $T/J > 0.3$. However, at lower temperatures, a discrepancy emerges, with a maximum error of approximately 10\% at zero temperature. We anticipate that increasing the accuracy of the calculations—by applying the decoupling approximation at higher-order equations of motion—could reduce this discrepancy and bring the results closer to those
from the Majorana-based numerical simulation.
Further improvements to the method will be addressed in future work.

\subsection{Excitation energies}
\begin{figure}[htbp]
  \includegraphics[width=1 \linewidth, angle=0]{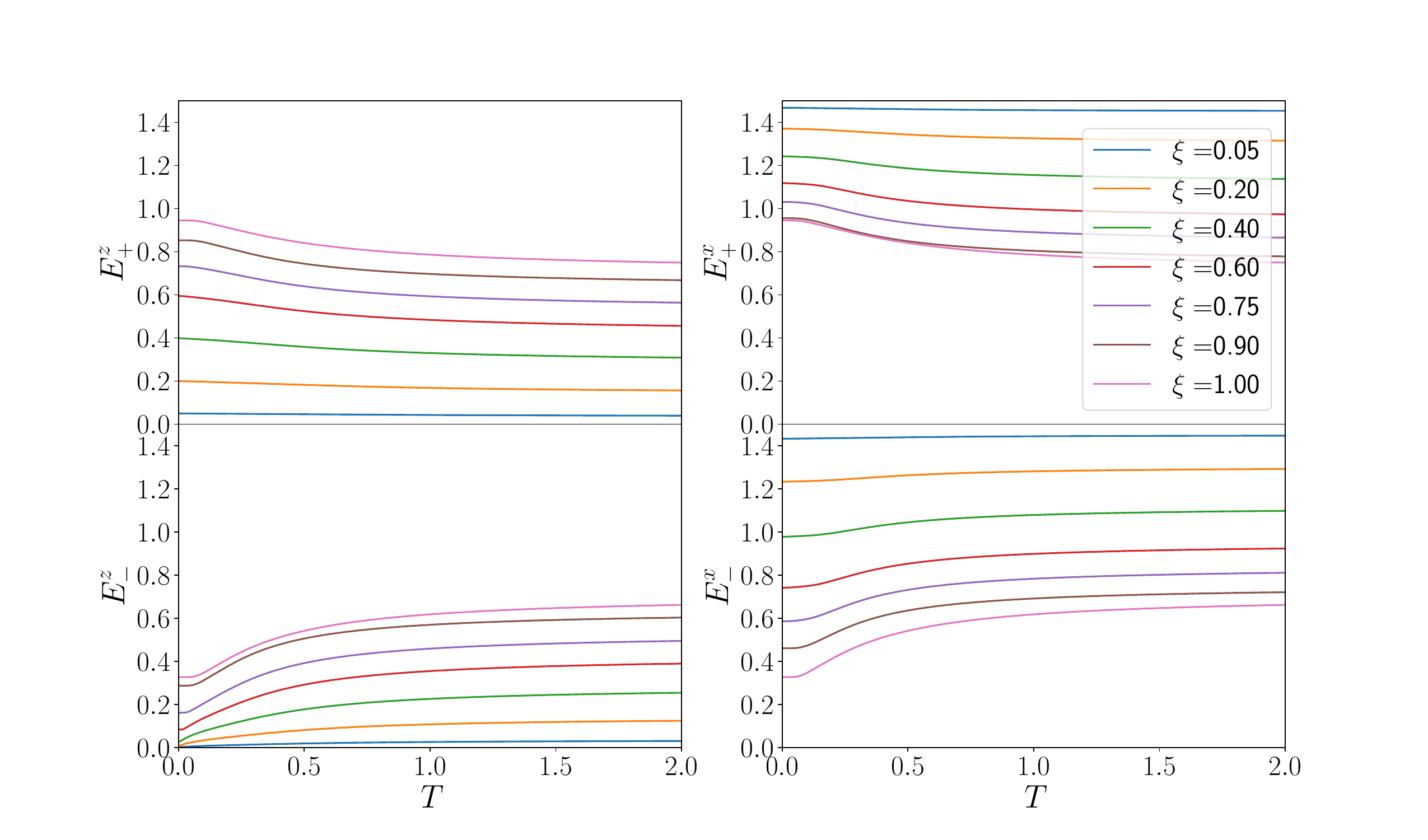}
  \caption{
    \label{fig:excitation-energies}
    Temperature dependence of the excitation energies in the Kitaev model. Two distinct excitation energies are shown, corresponding to excitations involving two lattice sites. The lower excitation energy is significantly higher than the vison excitation energy, as the Green's function primarily describes spin flip processes, which involve flipping two adjacent $\mathbb{Z}_2$ flux values, causing minor energy changes. The dominant contribution to the excitation energy comes from changes in the interaction energy between nearest-neighbor bonds.
  }
\end{figure}

The basic excitations of the system are encapsulated in the poles of the Green's functions, represented by the energies $E_{\pm}^z$ and $E_{\pm}^x$. In Fig.~\ref{fig:excitation-energies}, we show the temperature dependence of these excitation energies. Each excitation involves two lattice sites, leading to two distinct energy values. A natural interpretation is that $E_{-}^z$ and $E_{-}^x$ correspond to symmetric combinations of the two sites, while $E_{+}^z$ and $E_{+}^x$ correspond to anti-symmetric combinations.

Notably, the lower excitation energies are significantly higher than the energy of an adjacent $\mathbb{Z}_2$ vortex pair excitation, or vison pair excitation energy, $\Delta_v$. Analytical calculations based on Majorana fermions give $\Delta_v = 0.0658 J$ for the isotropic case \cite{Panigrahi2023}. This can be understood by considering that the Green's function describes not only the flipping of $\mathbb{Z}_2$ fluxes, but also simple spin-flip processes. A spin flip is typically accompanied by the flipping of two adjacent $\mathbb{Z}_2$ fluxes, but the energy cost of this process is minor. The dominant contribution to the excitation energy comes from changes in the interaction energy between nearest-neighbor bonds.

Let us explain how the application of a spin operator to any state induces the flipping of both adjacent $\mathbb{Z}_2$ fluxes and a single spin. As a concrete example, we focus on the seven spins labeled 0, 1, 2, 3, 7, 8, and 10 in Fig.~\ref{fig:honeycomb_lattice}. The $\mathbb{Z}_2$ flux operator $W$ for the hexagon consisting of sites labeled 0, 3, 8, 10, 7, and 2 is given by  
  \be
  W = \sigma _0^x\sigma _3^y\sigma _8^z\sigma _{10}^x\sigma _7^y\sigma _2^z.
  \ee
  Many states have an eigenvalue of $W$ equal to 1. Consider one such state,
  \be
  \left|  +  \right\rangle  = \frac{1}{{\sqrt 2 }}\left( {\left| { \uparrow  \uparrow  \uparrow  \uparrow  \downarrow  \uparrow } \right\rangle  + \left| { \downarrow  \downarrow  \uparrow  \downarrow  \uparrow  \uparrow } \right\rangle } \right),
  \ee
where the spin-up state is denoted by $\uparrow$ and the spin-down state by $\downarrow$. The ket-vectors on the right-hand side represent the spin states at sites 0, 3, 8, 10, 7, and 2, listed from left to right. It is straightforward to verify that  
$W\left|  +  \right\rangle  = \left|  +  \right\rangle$.
However, we observe that  
  \be
  W\sigma _0^y\left|  +  \right\rangle  =  - \sigma _0^y\left|  +  \right\rangle.
  \ee
  Thus, the $\mathbb{Z}_2$ flux value of $W$ is flipped by the application of $\sigma_0^y$ to $\left|  +  \right\rangle$.

Now, consider the spin-spin correlation functions. We find  
  \bea
  \left\langle  +  \right|\sigma _0^y\left( {\sigma _0^x\sigma _1^x} \right)\sigma _0^y\left|  +  \right\rangle
  &=&  - \left\langle  +  \right|\sigma _0^y\left( {\sigma _0^x\sigma _1^x} \right)\sigma _0^y\left|  +  \right\rangle, \\
  \left\langle  +  \right|\sigma _0^y\left( {\sigma _0^y\sigma _1^y} \right)\sigma _0^y\left|  +  \right\rangle  &=&  + \left\langle  +  \right|\sigma _0^y\left( {\sigma _0^y\sigma _1^y} \right)\sigma _0^y\left|  +  \right\rangle, \\
  \left\langle  +  \right|\sigma _0^y\left( {\sigma _0^z\sigma _1^z} \right)\sigma _0^y\left|  +  \right\rangle  &=&  - \left\langle  +  \right|\sigma _0^y\left( {\sigma _0^z\sigma _1^z} \right)\sigma _0^y\left|  +  \right\rangle.
  \eea
Therefore, certain spin-spin correlation functions change their signs. From this, we observe that the two-spin Green's function and spin-spin correlation functions involve the flipping of $\mathbb{Z}_2$ flux values and the sign change of the correlation functions.

For the isotropic case at zero temperature, the energy associated with this process is roughly estimated to be $0.24 J$ for a single $z$-bond or $x$-bond, as a spin flip changes $c_z$ to $-c_z$. Including contributions from other bonds, we expect a higher excitation energy. Indeed, for the isotropic case, we find $E_{-}^z = E_{-}^x = 0.327 J$ at zero temperature. This energy primarily originates from the simple spin-flip process.

Beyond the isotropic coupling, we observe that $E_-^z$ decreases as $\xi$ decreases. This can be understood by considering that, in this regime, the spins become more locally aligned. As a result, there are local "spin-wave" excitations, which require minimal energy to occur.

\subsection{Magnetic susceptibility}
\begin{figure}[htbp]
  \includegraphics[width=0.9 \linewidth, angle=0]{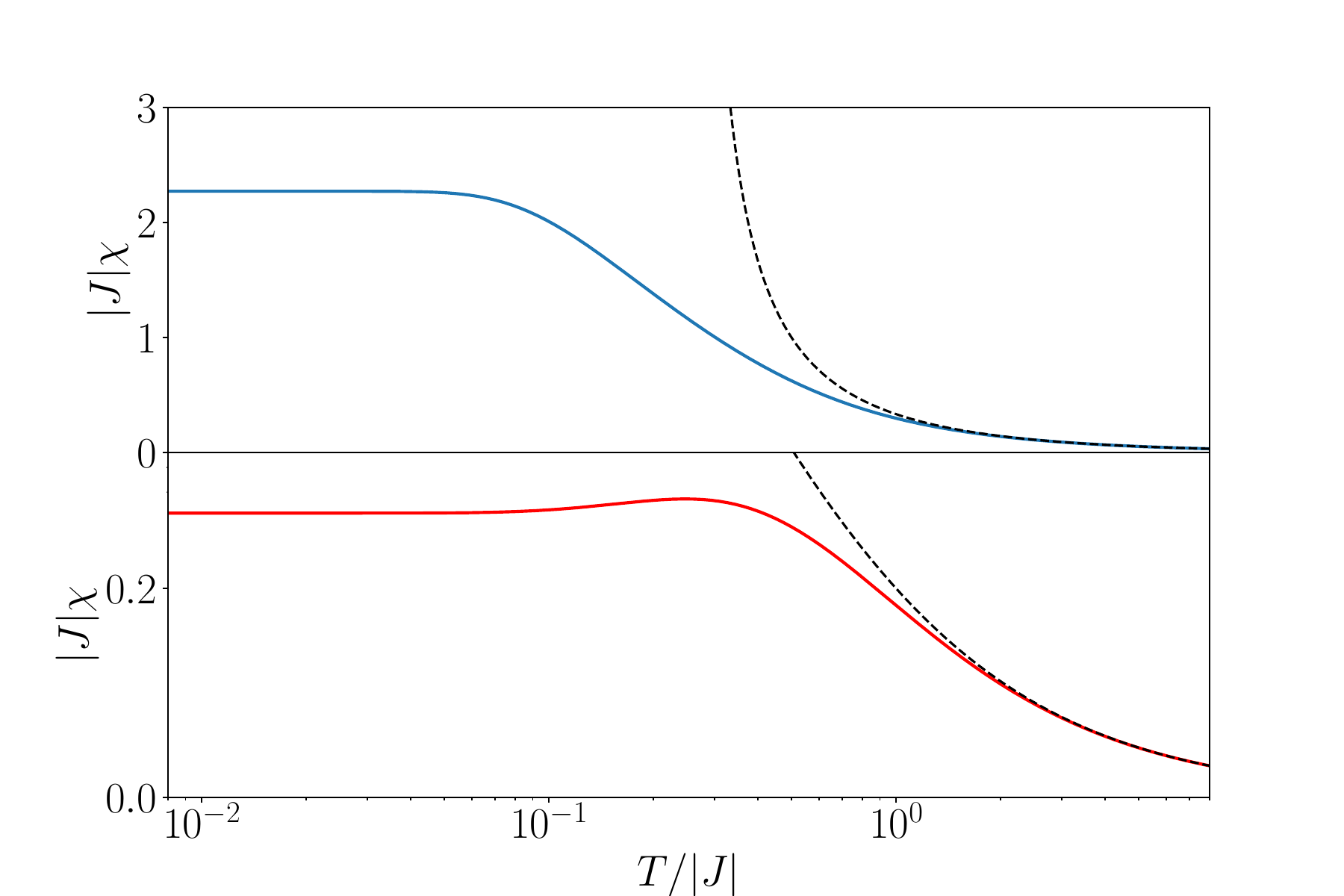}
  \caption{
    \label{fig:spin-susceptibility-isotropic}
    Temperature dependence of the spin susceptibility for the isotropic case. The upper panel shows the ferromagnetic case, while the lower panel represents the antiferromagnetic case. The dashed lines indicate the Curie-Weiss law for both cases.
  }
\end{figure}

One of the basic physical quantities accessible through experiments is the spin susceptibility. In Fig.~\ref{fig:spin-susceptibility-isotropic}, we present the temperature dependence of the spin susceptibility for the isotropic case. Both the ferromagnetic and antiferromagnetic cases follow the Curie-Weiss law at high temperatures. For the antiferromagnetic case, the magnitude of the spin susceptibility agrees with a Majorana-based numerical simulation\cite{Yoshitake2017b}.
However, for the ferromagnetic case, the calculated susceptibility is
approximately half of the Majorana-based numerical simulations result.

As the temperature decreases, the spin susceptibility in the ferromagnetic case increases and then saturates,
while the Majorana-based numerical simulation
predicts a peak around $T \simeq 0.02J$, which is absent in our calculations.
For the antiferromagnetic case, the overall behavior is consistent
with the Majorana-based numerical simulation\cite{Yoshitake2017b}, showing a broad peak around $T = O(0.1J)$. However, the Green's function approach predicts the peak position to be about three times higher than the Majorana-based numerical simulation result.
Notably, both ferromagnetic and antiferromagnetic cases exhibit temperature-independent paramagnetic behavior at low temperatures, reminiscent of the Pauli paramagnetism of itinerant fermions.

\begin{figure}[htbp]
  \includegraphics[width=0.9 \linewidth, angle=0]{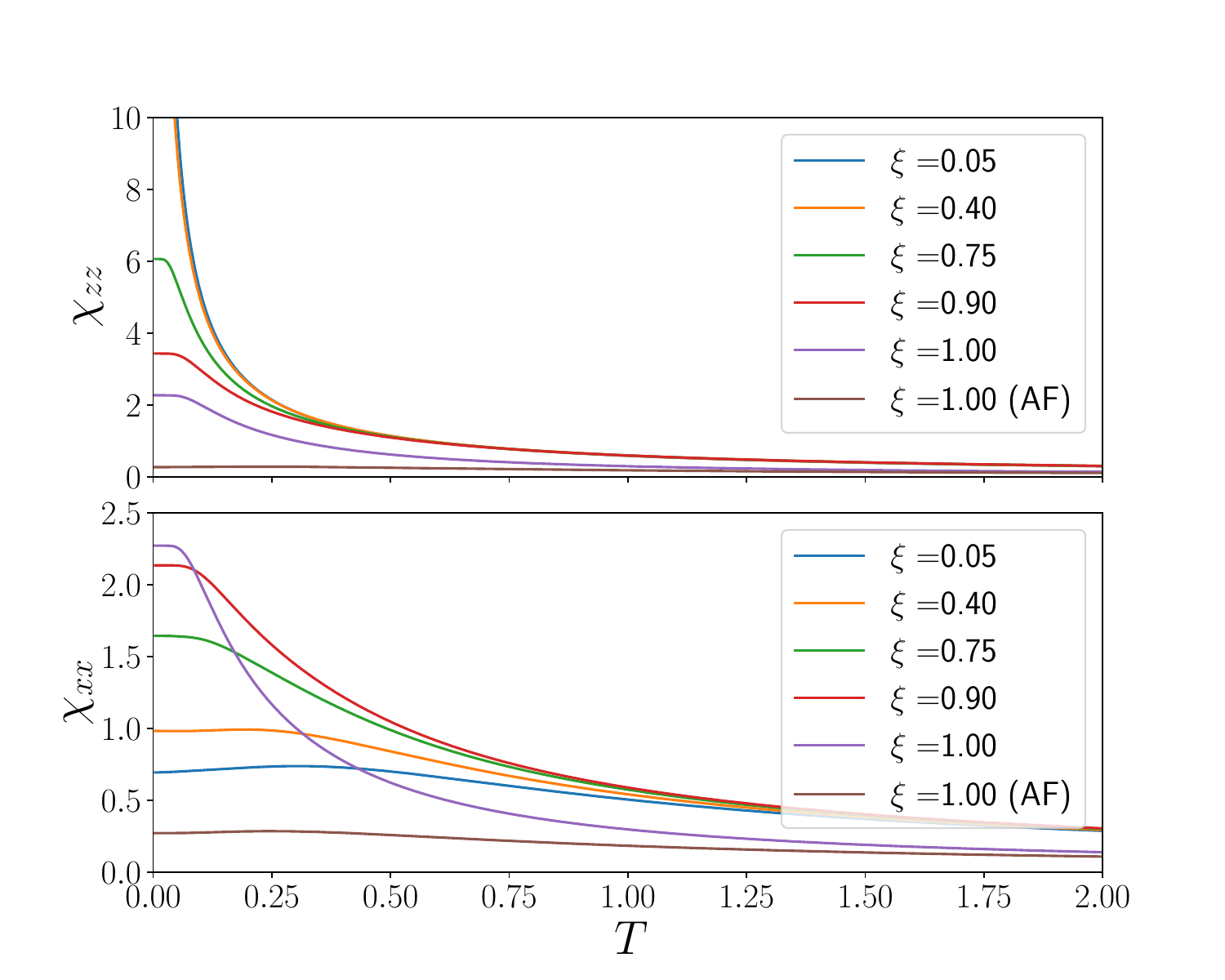}
  \caption{
    \label{fig:spin_susceptibility}
    Temperature dependence of the spin susceptibility in the Kitaev model. For $\xi \neq 1$, no difference is observed in the spin susceptibility between the ferromagnetic ($J_x > 0$ and $J_z > 0$) and antiferromagnetic ($J_x < 0$ and $J_z < 0$) cases.
  }
\end{figure}

We also compute the temperature dependence of the spin susceptibility for anisotropic cases, as shown in Fig.~\ref{fig:spin_susceptibility}. For $\xi \neq 1$, there is no difference between the ferromagnetic ($J_x > 0$ and $J_z > 0$) and antiferromagnetic ($J_x < 0$ and $J_z < 0$) cases, as noted in Ref.~\onlinecite{Kitaev2006}. Additionally, the difference between $\chi_{zz}$ and $\chi_{xx}$ in the anisotropic case can lead to an oscillation in the magnetic susceptibility when a magnetic field is applied in the plane and its direction is rotated. However, to accurately capture the anisotropy in the magnetic susceptibility, non-Kitaev interactions\cite{Chaloupka2010,Singh2012,Rau2014,Kim2016} must be considered.

In fact, the oscillating behavior observed in $\alpha$-RuCl$_3$ is associated with non-Kitaev interactions, as revealed by detailed experiments and theoretical analysis based on high-temperature expansions \cite{LampenKelley2018}.

\subsection{Dynamical structure factor}
We now investigate the spin dynamics, which is experimentally probed through inelastic neutron scattering. The cross section for such measurements is given by the dynamical structure factor, as defined in Eqs.~(\ref{eq:Sqw_each}) and (\ref{eq:Sqw}).

\begin{figure}[htbp]
  \includegraphics[width=0.9 \linewidth, angle=0]{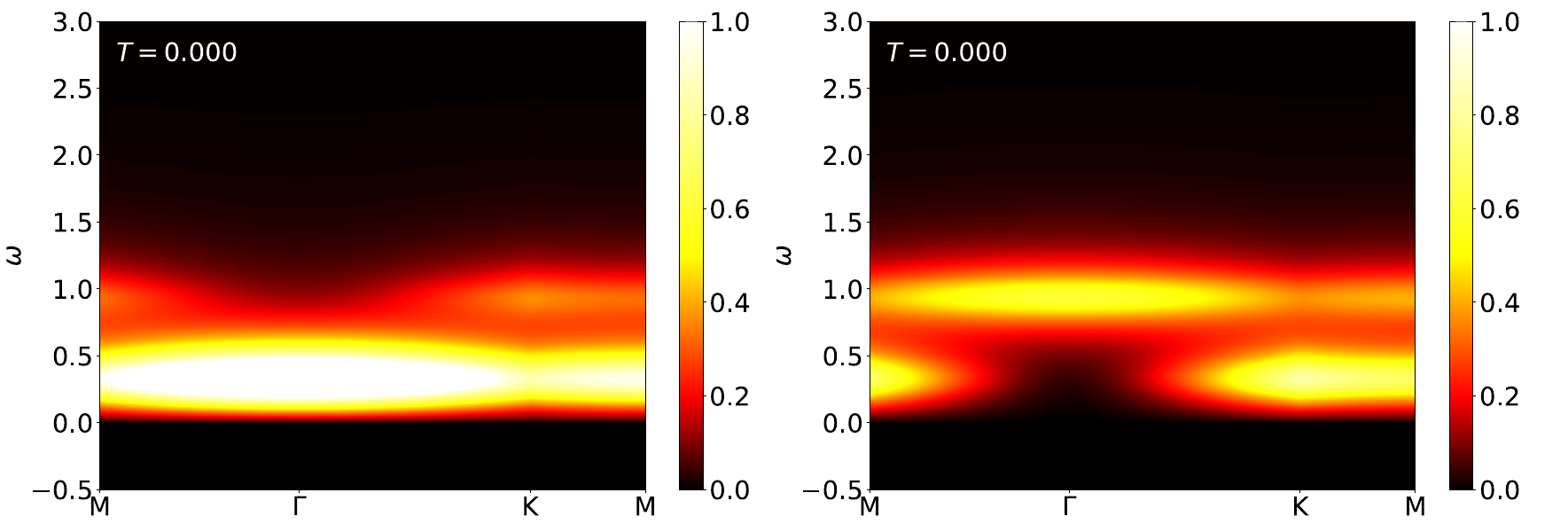}
  \caption{
    \label{fig:dsf_T0_iso}
Dynamical structure factor for the isotropic ferromagnetic case (left) and isotropic antiferromagnetic case (right) at zero temperature, plotted as a function of ${\bf q}$ and $\omega$ along the path M-$\Gamma$-K-M through the Brillouin zone. The calculations are performed with $\delta = 0.2$, and the energy is given in units of $|J|$.    
  }
\end{figure}

The results for the isotropic ferromagnetic and antiferromagnetic cases at zero temperature are shown in Fig.~\ref{fig:dsf_T0_iso}. Notably, the $\mathbf{q}$ dependence arises solely from the term $\cos \left( \mathbf{q} \cdot \mathbf{a}_\mu \right)$ in Eq.~(\ref{eq:Sqw_each}), as there is no dispersion in the excitation energies, $E_{\pm}^{x}$. The difference between the ferromagnetic and antiferromagnetic cases stems from a sign change in the numerator of $G_{0,1}^x(i\omega_n)$ (see Eq.~(\ref{eq:sol_G10_G01_iwn})). For the ferromagnetic case, $E_{-}^{x}$ contributes to $S(\mathbf{q}, \omega)$, while for the antiferromagnetic case, $E_{+}^{x}$ dominates. This difference manifests around the $\Gamma$ point for $\omega < 1$. Similar features have been observed in calculations based on Majorana fermions \cite{Knolle2014}, although there is no broad incoherent intensity in the high-energy region in our calculations.

\begin{figure}[htbp]
  \includegraphics[width=0.9 \linewidth, angle=0]{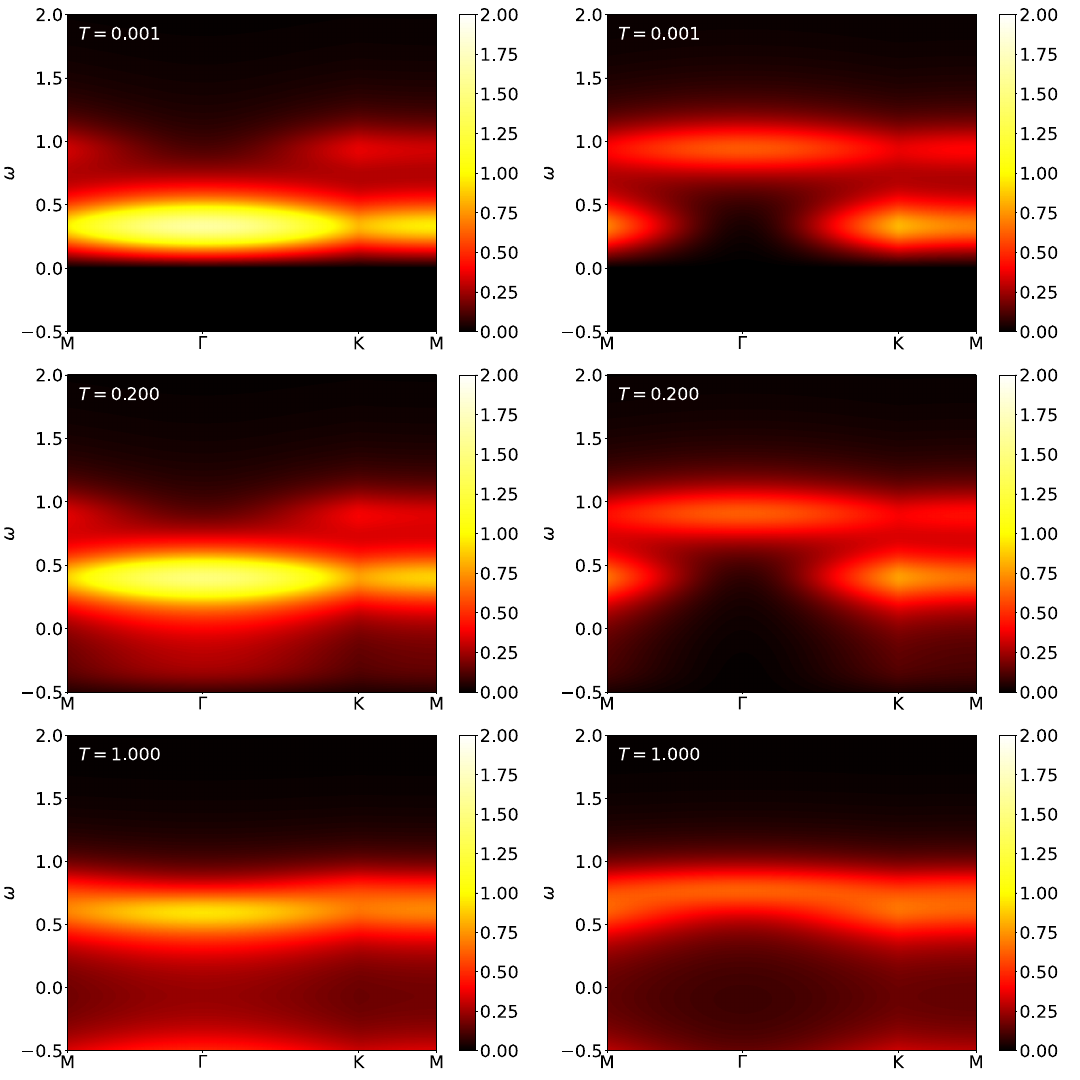}
  \caption{
    \label{fig:dsf_finite_T}
    Dynamical structure factor for the isotropic ferromagnetic case (left) and isotropic antiferromagnetic case (right) as a function of ${\bf q}$ and $\omega$ along the M-$\Gamma$-K-M path through the Brillouin zone at different temperatures. The calculations are performed with $\delta = 0.2$, with energy and temperature both in units of $|J|$. The overall features are consistent
    with a Majorana-based numerical simulation result\cite{Yoshitake2017b}.
  }
\end{figure}

We also compute $S(\mathbf{q}, \omega)$ for various temperatures in the isotropic case, as shown in Fig.~\ref{fig:dsf_finite_T}. As the temperature increases, the features associated with the excitations $E_{\pm}^{x}$ become increasingly smeared. Similar behavior is found
in the Majorana-based numerical simulation\cite{Yoshitake2017b}.
However, our calculations do not show any significant change around $T \sim 0.012 |J|$ in the width of the spectrum near the $\Gamma$ point, as there is no characteristic excitation with an energy of $\sim 0.012 |J|$.

%
%

\begin{figure}[htbp]
  \includegraphics[width=0.9 \linewidth, angle=0]{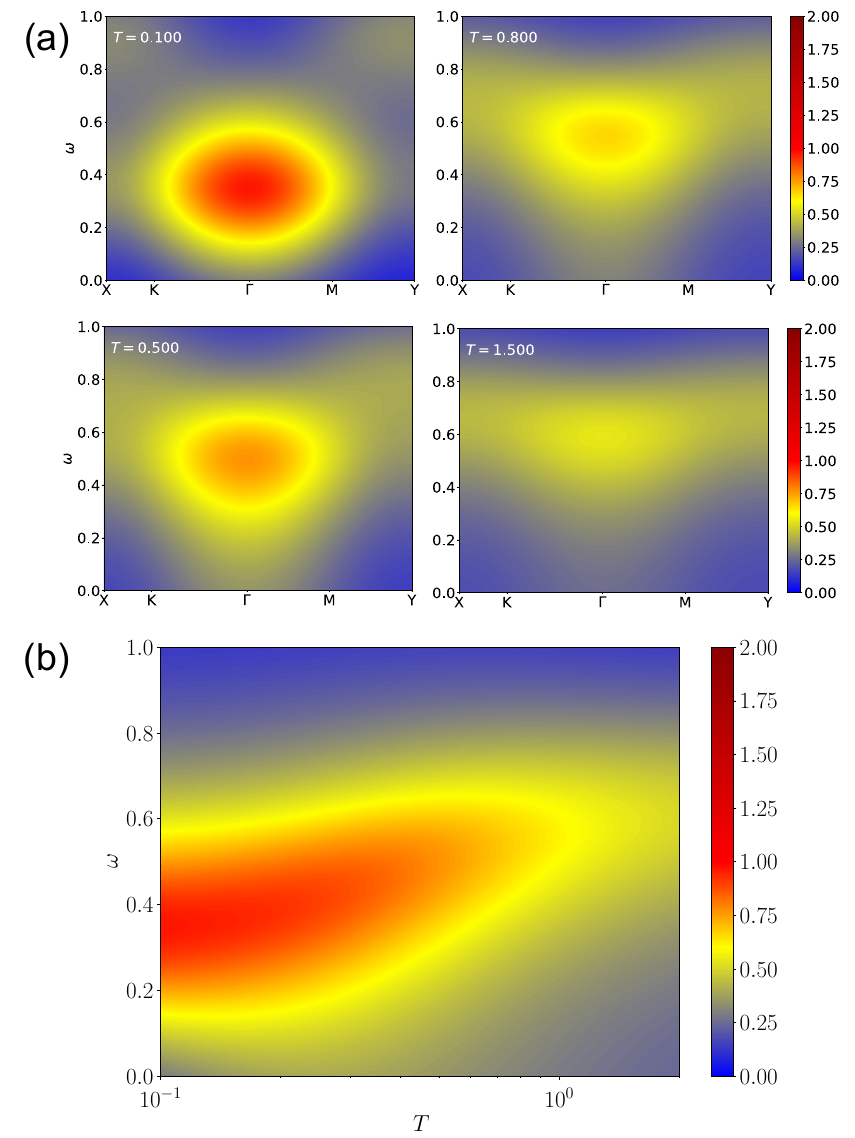}
  \caption{
    \label{fig:neutron}
    Temperature dependence of the dynamical spin structure factor as observed in inelastic neutron scattering. (a) At low temperatures, a clear excitation is visible around the $\Gamma$ point at $\omega/J \sim 0.3$. As the temperature increases, the excitation shifts toward higher energies, and an hourglass-like continuum emerges. This behavior is qualitatively consistent with experimental findings \cite{Do2017}. In the calculations, the Bose-factor correction is applied.
    (b) 
      Heatmap plot of the dynamical spin structure factor in the $\omega$-$T$  plane at the $\Gamma$ point. The results are consistent with those obtained from Majorana-based numerical simulations \cite{Do2017}, except in the low-energy region ($\omega < 0.2$), which is not accessible via neutron scattering experiments \cite{Do2017}.
  }
\end{figure}

To compare with inelastic neutron scattering results, we present data for different temperatures in Fig.~\ref{fig:neutron}. As shown in Fig.~\ref{fig:neutron}(a), a clear excitation appears around the $\Gamma$ point at approximately $\omega/J \sim 0.3$ at low temperatures. As the temperature increases, the peak gradually shifts toward higher energies, and an hourglass-like feature develops. This behavior is qualitatively consistent with experimental observations \cite{Do2017}.
  Figure~\ref{fig:neutron}(b) presents a heatmap plot of the dynamical spin structure factor as a function of $\omega$ and $T$ at the $\Gamma$ point, illustrating the temperature dependence of spin excitations across a wide energy range.
  Our results closely align with Majorana-based numerical simulations\cite{Do2017},
  capturing the essential features of the spin dynamics.
  However, notable discrepancies are observed in the low-energy region ($\omega < 0.2$),
  which was inaccessible to neutron scattering experiments\cite{Do2017}.

\subsection{Nuclear magnetic resonance}
As another physical quantity, we consider the longitudinal relaxation rate, $1/T_1$, observed in nuclear magnetic resonance experiments. The hyperfine coupling constant varies depending on the material and is influenced by the wave vector characteristic of the spin correlations. To isolate the components that are independent of the hyperfine coupling, we analyze the contributions to $1/T_1$ separately. A similar decomposition is performed in Majorana-based numerical simulations \cite{Yoshitake2017b}. In the Kitaev model, there are only two terms that contribute to $1/T_1$: one from the on-site component
\be
\frac{1}{T_1^{00}}
= S_{0,0}^{xx}\left( {\omega  = {\omega _0}} \right)
+ S_{0,0}^{yy}\left( {\omega  = {\omega _0}} \right),
\ee
and the other from the nearest-neighbor site component
\be
\frac{1}{T_1^{{\rm{NN}}}}
 = S_{0,1}^{xx}\left( {\omega  = {\omega _0}} \right)
+ S_{0,2}^{yy}\left( {\omega  = {\omega _0}} \right).
\ee
From Eq.~(\ref{eq:Sqw_each}), by setting ${\bf q}=0$, we obtain
\be
S_{00}^{\mu \mu }\left( \omega  \right)
=  - \frac{1}{\pi }\frac{1}{{1 - {e^{ - \beta \omega }}}}
{\mathop{\rm Im}\nolimits} G_{0,0}^\mu \left( {i{\omega _n} \to \omega  + i\delta } \right),
\ee
and 
\be
S_{0\mu}^{\mu \mu }\left( \omega  \right)
=  - \frac{1}{\pi }\frac{1}{{1 - {e^{ - \beta \omega }}}}
{\mathop{\rm Im}\nolimits} G_{0,{i_\mu }}^\mu
\left( {i{\omega _n} \to \omega  + i\delta } \right).
\ee

\begin{figure}[htbp]
  \includegraphics[width=0.9 \linewidth, angle=0]{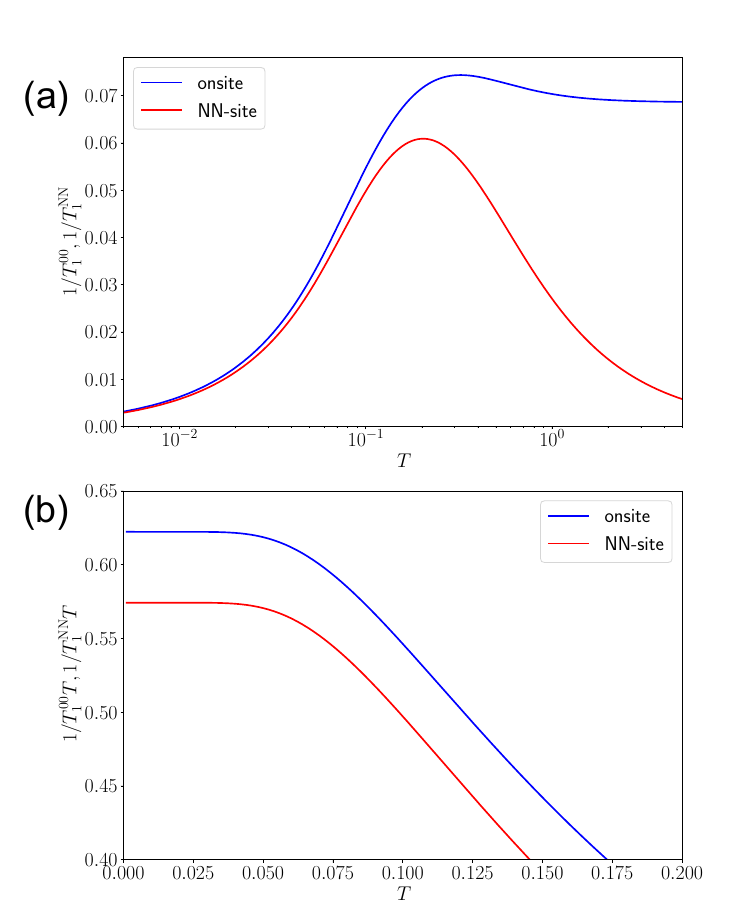}
  \caption{
    \label{fig:T1-inv}
    (a) Temperature dependence of the longitudinal relaxation rates $1/T_1^{00}$ (on-site component) and $1/T_1^{\rm{NN}}$ (nearest-neighbor component) for the isotropic ferromagnetic case. At high temperatures, $1/T_1^{00}$ remains almost temperature-independent, while $1/T_1^{\rm{NN}}$ exhibits a monotonic decrease. As the temperature decreases, $1/T_1^{00}$ shows a hump, and both components rapidly decrease at low temperatures. (b) Temperature dependence of $1/T_1 T$ for the same components. Both approach constant values as $T \rightarrow 0$, suggesting fermionic behavior, which aligns with the temperature-independent paramagnetism observed at low temperatures. In both figures, we take $\omega_0 = 10^{-4}$ and $\delta = 0.3$.
  }
\end{figure}

Figure~\ref{fig:T1-inv}(a) shows the temperature dependence of $1/T_1^{00}$ and $1/T_1^{\rm{NN}}$ for the isotropic ferromagnetic case. The on-site component, $1/T_1^{00}$, remains almost temperature-independent at high temperatures. As the temperature decreases, it exhibits a hump before rapidly decreasing. The nearest-neighbor component, $1/T_1^{\rm{NN}}$, displays a peak around $T = 0.2J$. At low temperatures, $1/T_1^{\rm{NN}}$ decreases rapidly, similar to $1/T_1^{00}$, while at high temperatures, $1/T_1^{\rm{NN}}$ shows a monotonic decrease, contrary to the behavior of $1/T_1^{00}$.

Similar behavior at high temperatures has been observed in a Majorana-based numerical simulation \cite{Yoshitake2017b}. However, in these simulations, both $1/T_1^{00}$ and $1/T_1^{\rm{NN}}$ exhibit a similar peak structure, with the peak located at $T \simeq 0.04J$. It has been suggested \cite{Motome2020} that an activation behavior can be seen at low temperatures, where the energy gap is associated with $\mathbb{Z}_2$ flux excitations. However, our data reveals a different behavior, as shown in Fig.~\ref{fig:T1-inv}(b), where we plot the temperature dependence of $1/T_1 T$. At low temperatures, both components approach constant values as $T \rightarrow 0$, suggesting fermionic behavior, which is consistent with the temperature-independent paramagnetic behavior observed earlier.

\section{Summary and Discussion}
\label{sec:summary}
In this work, we investigated the Kitaev model using the equation of motion for the spin Green's function. The formalism was developed up to the second order of the equation of motion, followed by the application of the decoupling approximation. We then solved the resulting self-consistent equations for the correlation functions, enabling the computation of various physical quantities,
including spin susceptibility,
dynamical spin structure factor,
and the longitudinal relaxation rate.

Our results were primarily compared with numerical simulations based on the Majorana fermion approach. While there are notable similarities between the two methods, especially at high temperatures, important differences arise. For example, in the case of spin susceptibility, both ferromagnetic and antiferromagnetic cases follow the Curie-Weiss law at high temperatures, with the antiferromagnetic susceptibility agreeing well
with the Majorana-based numerical simulation.
However, for the ferromagnetic case, our result is approximately half
of the Majorana-based numerical simulation, and the low-temperature peak observed
is absent in our calculations. Furthermore, the Green's function approach revealed temperature-independent paramagnetic behavior at low temperatures, reminiscent of Pauli paramagnetism, which differs from the behavior predicted by Majorana-based simulations.

We also studied the dynamical spin structure factor, which exhibited features similar to those seen in Majorana fermion studies, such as $\Gamma$-point excitations in both ferromagnetic and antiferromagnetic cases. However, our calculations did not capture the broad incoherent intensity at high energy seen in the Majorana fermion-based results. The temperature dependence of the dynamical structure factor in our calculations revealed smearing of excitation features as temperature increased, consistent with previous findings.

For the longitudinal relaxation rate $1/T_1$, our results show distinct temperature-dependent behaviors for both the on-site and nearest-neighbor components. At low temperatures, both components approach constant values, suggesting fermionic behavior. In contrast, Majorana-based numerical simulations predict activation behavior associated with $\mathbb{Z}_2$ flux excitations, leading to differing interpretations of the low-temperature regime.

While the Green's function approach performs well at high temperatures, the discrepancies observed at lower temperatures suggest that further improvements could be achieved by extending the decoupling approximation to higher orders in the equation of motion. These extensions, along with the exploration of non-Kitaev interactions, may help bridge the gap between the Green's function method and Majorana-based approaches in future studies.

It is important to note that, in order to compute the thermal Hall conductivity within our formalism, two key extensions are necessary. First, we must include the effect of an external magnetic field, as it plays a crucial role in driving the Kitaev model into different physical regimes, particularly the gapped chiral Kitaev spin liquid. This is exemplified by the recent experimental observation of half-quantized thermal Hall conductivity in the material $\alpha$-RuCl$_3$ under an external magnetic field \cite{Kasahara2018}. This state is characterized by a chiral Majorana edge mode, and the half-quantized thermal Hall conductivity is considered a unique signature of this Majorana edge state \cite{Kitaev2006}. Incorporating a magnetic field is critical for exploring the connection between Majorana fermions and topological excitations in spin liquids.
  Second, in order to compute certain quantities, we need to calculate multi-spin Green's functions. For instance, four-spin Green's functions are required to compute the specific heat because this involves calculating the expectation value of the square of the Hamiltonian. However, an approximate calculation can also be performed by taking the temperature derivative of the internal energy. For thermal transport, we need to compute the correlation function of energy currents. Since the energy current consists of three-spin operators, as derived from the general formula \cite{Lee2015thermal}, calculating thermal transport requires six-spin Green's functions.

  Additionally, introducing a magnetic field complicates the analysis because it invalidates the property that certain Green's functions and correlation functions are identically zero in the zero-field Kitaev model. This adds complexity to the calculation, requiring a new formalism capable of handling the effects of the magnetic field on the system. We anticipate that such a formalism would provide deeper insights into the behavior of Majorana fermions, especially since the introduction of a magnetic field is expected to alter the localized nature of spin correlations and potentially reveal novel collective excitations.

  Although the computation of thermal conductivity and specific heat is left for future work, our formalism is well-equipped to address these quantities with suitable modifications. Expanding the current approach to include four-spin Green's functions, magnetic fields, and non-Kitaev interactions will enable a more comprehensive exploration of the thermal and topological properties of the Kitaev model, particularly in regimes pertinent to recent experimental observations.

\appendix
\section{Vanishing Green's functions and correlation functions}
\label{app:exact-results}
Due to the special properties of the Kitaev model, it can be rigorously shown that certain Green's functions and correlation functions vanish \cite{Baskaran2007, Motome2020}. Below, we briefly explain the underlying mechanism.

\begin{figure}[htbp]
  \includegraphics[width=0.5 \linewidth, angle=0]{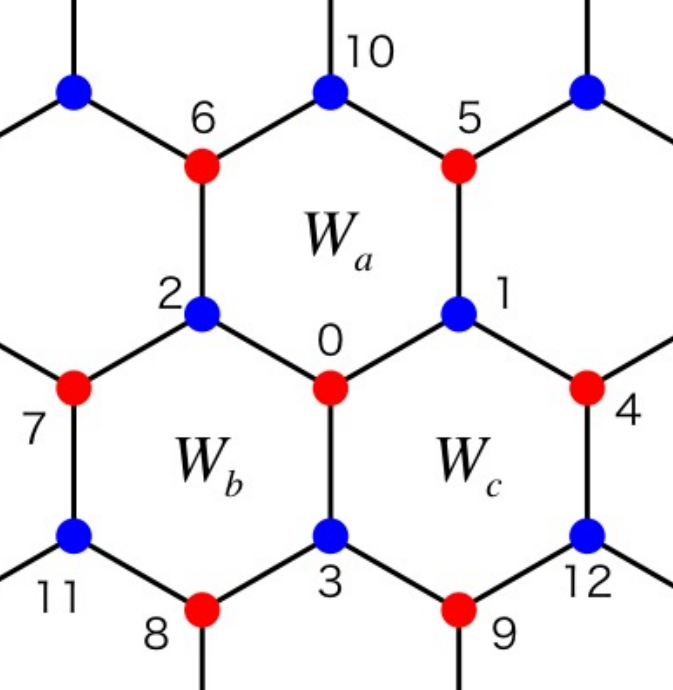}
  \caption{
    \label{fig:Z2_fluxes}
     Illustration of the Kitaev model's honeycomb lattice showing the three $\mathbb{Z}_2$ flux operators, $W_a$, $W_b$, and $W_c$. These flux operators are associated with the hexagons on the lattice and play a crucial role in determining the behavior of Green's functions and correlation functions in the model.
  }
\end{figure}

Consider three $\mathbb{Z}_2$ flux operators, $W_a$, $W_b$, and $W_c$, as shown in Fig.~\ref{fig:Z2_fluxes}.
In terms of the Pauli matrices at each site, the explicit definitions are given by
\bea
   {W_a} &=& \sigma _0^z\sigma _2^x\sigma _6^y\sigma _{10}^z\sigma _5^x\sigma _1^y,\\
   {W_b} &=& \sigma _0^x\sigma _3^y\sigma _8^z\sigma _{11}^x\sigma _7^y\sigma _2^z,\\
   {W_c} &=& \sigma _0^y\sigma _1^z\sigma _4^x\sigma _{12}^y\sigma _9^z\sigma _3^x.
   \eea
Since these $\mathbb{Z}_2$ flux operators commute with the Hamiltonian ${\cal H}$, they can be simultaneously diagonalized. Let $\left| n \right\rangle$ represent the eigenstate of ${\cal H}$, $W_a$, $W_b$, and $W_c$, such that
   \bea
   {\cal H} \left| n \right\rangle  &=& {E_n}\left| n \right\rangle, \\
   {W_a}\left| n \right\rangle  &=& {w_a}\left| n \right\rangle, \\
   {W_b}\left| n \right\rangle  &=& {w_b}\left| n \right\rangle, \\
   {W_c}\left| n \right\rangle  &=& {w_c}\left| n \right\rangle,
   \eea
where $w_a$, $w_b$, and $w_c$ are the eigenvalues corresponding to $W_a$, $W_b$, and $W_c$, respectively. Since $W_a^2 = 1$, the eigenvalues $w_a$, $w_b$, and $w_c$ take values $\pm 1$. 

It is important to note that ${W_a}\sigma _0^x = - \sigma _0^x{W_a}$, which leads to
   \be
      {W_a}\sigma _0^x\left| n \right\rangle  =  - {w_a}\sigma _0^x\left| n \right\rangle.
      \ee
Similarly, we find that
      \bea
      {W_b}\sigma _0^x\left| n \right\rangle  &=&  + {w_b}\sigma _0^x\left| n \right\rangle,\\
      {W_c}\sigma _0^x\left| n \right\rangle  &=&  - {w_c}\sigma _0^x\left| n \right\rangle.
      \eea
Thus, the $\mathbb{Z}_2$ flux values on the adjacent hexagons that share the $x$ bond emanating from the 0-site change their signs, while the $\mathbb{Z}_2$ flux values associated with the other hexagons remain unchanged. In general, the $\gamma$ component of the Pauli matrix at site $j$ flips the signs of the pair of $\mathbb{Z}_2$ flux values defined on the hexagons sharing the $\gamma$ bond that emanates from site $j$.

Now consider the Green's function
      \be
      G_{i,j}^{\mu \nu }\left( \tau  \right) =  - \left\langle {{T_\tau }S_i^\mu \left( \tau  \right)S_j^\nu \left( 0 \right)} \right\rangle.
      \ee
For $\tau > 0$, we have
      \be
      G_{i,j}^{\mu \nu }\left( \tau  \right) =  - \frac{1}{{4Z}}\sum\limits_n {\left\langle n \right|{e^{\tau {\cal H}}}\sigma _i^\mu {e^{ - \tau {\cal H}}}\sigma _j^\nu \left| n \right\rangle },
      \ee
where $Z$ is the partition function. The Pauli matrices $\sigma _i^\mu$ and $\sigma _j^\nu$ flip the $\mathbb{Z}_2$ flux values, as explained above, while the Hamiltonian ${\cal H}$ does not. Therefore, if the $\mathbb{Z}_2$ flux values flipped by $\sigma _j^\nu$ acting on $\left| n \right\rangle$ are not flipped back by $\sigma _i^\mu$, the corresponding terms do not contribute to $G_{i,j}^{\mu \nu }(\tau)$. 

For example, this explains why $G_{0,1}^{xy}(\tau) = 0$ and $G_{0,2}^{xx}(\tau) = 0$. Similarly, by setting $\tau = 0$, we find that $\left\langle \sigma _0^x\sigma _1^y \right\rangle = 0$ and $\left\langle \sigma _0^x\sigma _2^x \right\rangle = 0$, among others.

\begin{acknowledgments}
  The authors thank D. Sasamoto for helpful discussions and valuable comments.
\end{acknowledgments}


\bibliography{../../../../refs/lib}
\end{document}